\documentclass[preprint,prd,superscriptaddress,nofootinbib,floatfix]{revtex4}

\usepackage{graphicx}
\usepackage{bm}
\usepackage{multirow}

\begin{document}

\preprint{OU-HET-689-2010}
\preprint{UTHEP-617}
\preprint{KEK-CP-249}

\title{
  Determination of the chiral condensate from QCD Dirac spectrum on the lattice
}

\newcommand{\Taiwan}{
Physics Department and Center for Quantum Science and Engineering,
National Taiwan University, Taipei 10617, Taiwan
}

\newcommand{\Tsukuba}{
  Graduate School of Pure and Applied Sciences,
University of Tsukuba, Tsukuba, Ibaraki 305-8571, Japan
}

\newcommand{\CCS}{
Center for Computational Sciences,
University of Tsukuba, Tsukuba, Ibaraki 305-8577, Japan
}
\newcommand{\KEK}{
  KEK Theory Center,
  High Energy Accelerator Research Organization (KEK),
  Tsukuba 305-0801, Japan
}
\newcommand{\GUAS}{
  School of High Energy Accelerator Science,
  The Graduate University for Advanced Studies (Sokendai),
  Tsukuba 305-0801, Japan
}
\newcommand{\YITP}{
  Yukawa Institute for Theoretical Physics, 
  Kyoto University, Kyoto 606-8502, Japan
}
\newcommand{\Osaka}{
  Department of Physics, Osaka University,
  Toyonaka 560-0043, Japan
}

\author{H.~Fukaya}
\affiliation{\Osaka}

\author{S.~Aoki}
\affiliation{\Tsukuba}
\affiliation{\CCS}

\author{T.W.~Chiu}
\affiliation{\Taiwan}

\author{S.~Hashimoto}
\affiliation{\KEK}
\affiliation{\GUAS}

\author{T.~Kaneko}
\affiliation{\KEK}
\affiliation{\GUAS}

\author{J.~Noaki}
\affiliation{\KEK}

\author{T.~Onogi}
\affiliation{\Osaka}

\author{N.~Yamada}
\affiliation{\KEK}
\affiliation{\GUAS}

\collaboration{JLQCD and TWQCD collaborations}
\noaffiliation


\begin{abstract} 
  We calculate the chiral condensate of QCD with 
  2, 2+1 and 3 flavors of sea quarks.
  Lattice QCD simulations are performed employing dynamical overlap fermions
  with up and down quark masses covering a range between 
  3 and 100~MeV.
  On $L\sim 1.8$--1.9 fm  lattices 
  at a lattice spacing $\sim$ 0.11~fm, we calculate the
  eigenvalue spectrum of the overlap-Dirac operator.
  By matching the lattice data 
  with the analytical prediction from chiral perturbation theory
  at the next-to-leading order, 
  the chiral condensate in the massless limit
  of up and down quarks 
  is determined. 
\end{abstract}
\maketitle

\section{Introduction}
\label{sec:intro}

Chiral condensate $\Sigma$ in Quantum Chromodynamics (QCD) is not a
directly accessible quantity in experiment, 
yet plays a crucial role in the low-energy dynamics of QCD as 
an order parameter of chiral symmetry breaking.
When $\Sigma$ in the limit of massless quarks is nonzero, 
chiral symmetry is spontaneously broken and hadrons acquire a mass of
order $\Lambda_{\mathrm{QCD}}$, the QCD scale.
Only the pion remains massless as the Nambu-Goldstone boson;
its dynamics is well described by an effective theory known as
chiral perturbation theory (ChPT) \cite{Weinberg:1968de,Gasser:1983yg}.
$\Sigma$ is one of two most fundamental parameters
in ChPT (the other is the pion decay constant $F$) and appears in the
predictions of various physical quantities. 

Calculation of $\Sigma$, the expectation value of the scalar density
operator $\Sigma=-\langle\bar{q}q\rangle$, 
from the first principles of QCD requires nonperturbative
techniques. 
In this paper we report on a numerical calculation of $\Sigma$ in
lattice QCD including the effects of up, down and strange sea quarks.
We investigate the low-lying eigenvalue spectrum $\rho(\lambda)$ of
the Dirac operator, which is related to $\Sigma$ through the
Banks-Casher relation $\rho(0)\to\Sigma/\pi$ \cite{Banks:1979yr}
and its extension to the case of nonzero $\lambda$.
Since only the low-lying eigenvalues are relevant, one can avoid
contamination from ultraviolet divergence of the scalar density operator
$\bar{q}q$, which is of order $m/a^2$ at a finite quark mass
$m$ and a lattice spacing $a$.


The Banks-Casher relation is satisfied in the limit of massless quarks
after taking the large volume limit (the thermodynamical limit), which
is the meaning of the arrow in the relation $\rho(0)\to\Sigma/\pi$. 
When the massless limit is taken at a finite volume, the vacuum
expectation value of $\bar{q}q$ shows a critical fluctuation which
leads to a divergent correlation length and vanishing chiral condensate.
Taking the thermodynamical limit, on the other hand, is numerically
expensive in practical lattice calculations.
In this study, we use the low energy effective theory as a guidance
of volume and quark mass scalings.
Namely, once the scaling behavior predicted by the effective theory is
confirmed by the lattice data, the infinite volume and chiral limit
according to the scaling can be safely taken.
The scaling we consider is that for varying the eigenvalue $\lambda$, 
volume $V$ and the quark mass $m$. 

We use the ChPT formula for the low-lying eigenvalue spectrum of the
Dirac operator \cite{Damgaard:2008zs}, which is valid in both the $p$ 
and $\epsilon$ regimes. 
The $\epsilon$ regime is a parameter region where the quark mass is
so small that the Compton wavelength of the pion is longer than the
extent of the finite-volume space-time.
In this regime, the constant mode of the pion field has to be
integrated out over the group manifold
when the path integral is evaluated, which is a nonperturbative
calculation in the sense that one does not use the expansion
in small pion field.
At the leading order of the so-called $\epsilon$ expansion, the system
is equivalently described by the chiral Random Matrix Theory (ChRMT),
with which a number of theoretical predictions for the low-lying Dirac
spectrum have been derived
\cite{Damgaard:1997ye,Wilke:1997gf,Akemann:1998ta,Damgaard:2000ah}.
In the other regime ($p$ regime), where the pion Compton wavelength
fits in the volume, the conventional ChPT applies and the calculation
of the Dirac operator spectrum is available as well
\cite{Smilga:1993in,Osborn:1998qb}.
The new method given in \cite{Damgaard:2008zs} consistently combines
the both results within a systematic expansion, and thus is valid in both
regimes as well as in between. 
By lattice calculations, we produce the data at various sets of quark masses
to firmly test this analytic expectation.

Application of the ChPT formula for the low-lying Dirac spectrum
requires a good control of the chiral symmetry in the calculation.
In this work, we use the overlap-Dirac operator
\cite{Neuberger:1997fp,Neuberger:1998wv} which satisfies the
Ginsparg-Wilson relation \cite {Ginsparg:1981bj} 
and thus realizes a {\it modified} chiral symmetry on the lattice
\cite{Luscher:1998pqa}.
Since the chiral effective theory is constructed only assuming the
presence of chiral symmetry, the same construction as in the
continuum theory can be applied to lattice QCD with overlap fermions.
Although the overlap-Dirac eigenvalues lie on
a circle on the complex plane, 
the physical 
imaginary part is uniquely identified up to $O(a^2)$ discretization 
effects. 

In our previous works \cite{Fukaya:2007fb,Fukaya:2007yv}
we performed large-scale lattice simulations of two-flavor QCD
using the overlap fermion formulation 
\cite{Aoki:2008tq} and calculated low-lying Dirac eigenvalues.
By matching the lowest eigenvalue spectrum with the ChRMT
expectations, we extracted $\Sigma$.
Since the ChRMT corresponds to the leading order of the
$\epsilon$ expansion in ChPT, this result is subject to NLO or $O(1/F^2V^{1/2})$
corrections, which could be sizable on the lattice of size 
$V=L^3 T\sim (1.9\;\mbox{fm})^4$ used in that study.
Another limitation was that the quark mass must be very small to apply
the $\epsilon$ regime formula, and the runs in the $p$ regime could
not be used in the analysis.

Application of the new formula \cite{Damgaard:2008zs} is attempted for
the first time in our recent work \cite{Fukaya:2009fh,Fukaya:2010ih}
in which next-to-leading order (NLO) corrections are
included in the analysis.
Using the 2+1-flavor QCD data generated with dynamical overlap
fermions on $16^3\times 48$ lattices, the value of $\Sigma$ in the
chiral limit of two light quarks is obtained.
The present paper provides an extensive description of this work. 

In this paper, we analyze the low-lying Dirac spectrum mainly on the 2+1-flavor QCD
simulations, where the strange quark mass $m_s$ is fixed near its
physical value.
The mass of degenerate up and down quarks, $m_{ud}$, covers a range between $m_s/5$ and 
$m_s$ in the $p$ regime lattice ensembles, which are generated for
calculations of various physical quantities including the
pion mass and decay constant \cite{JLQCD:2009sk}.
We also generate an $\epsilon$ regime ensemble, where the up and down
quarks are kept nearly massless while $m_s$ is fixed near the
physical value.
The NLO ChPT formula allows us to combine these data
to obtain $\Sigma$ in the chiral limit of up and down quarks.
We also extract the pion decay constant $F$ and one of the NLO
low energy constants (LECs) $L_6$ from the correction terms.

As demonstrated in the following sections, the ChPT formula provides
information on the shape of the low-lying Dirac spectrum, with
which we can test the agreement between the formula and the lattice
data in detail. 
The volume dependence gives a critical test, since it is 
essentially controlled by $\Sigma$ and $F$.
At some parameter points, we compare the data obtained on a larger
($24^3\times 48$) lattice to those on the smaller ($16^3\times 48$) volume
to check if the ChPT prediction consistently describes the difference.
The sea quark mass dependence of the chiral condensate is partly controlled by
$L_6$ but we also expect a nonanalytic dependence due to the
pion-loop effects that should be observed in the lattice data.
These nontrivial consistency checks are performed to gain
confidence in the final result for $\Sigma$. 
 
In addition to the main analysis in 2+1-flavor QCD, 
we also investigate two-flavor QCD and three-flavor QCD 
where $m_{ud}$ and $m_s$ are degenerate.
For the case of two-flavor QCD, the lattice data are the same as in our
previous studies \cite{Fukaya:2007fb,Fukaya:2007yv}, but we use
the ChPT formula valid at NLO in this new analysis. 
The degenerate three-flavor QCD configurations are newly generated for
this study at two light quark masses.
We thus obtain the chiral condensate $\Sigma$ for these variants of
QCD.


This paper is organized as follows.
In Section~\ref{sec:ChPT}, 
we briefly explain how the chiral condensate
is determined from the Dirac eigenvalue density
and how the finite volume effects are removed using ChPT.
Some technical aspects of lattice QCD simulations are described in 
Section~\ref{sec:lattice} and the numerical results are discussed in
the following sections.  
First, we describe the strategy to extract the LECs
from lattice data of the spectral density 
in Section~\ref{sec:extraction}.
Second, numerical scaling tests of the NLO ChPT formula are
presented in Section~\ref{sec:scaling}.
We then proceed to the determination of $\Sigma$ in the chiral limit
in Section~\ref{sec:determination}. 
Summary of this analysis and conclusions are given in
Section~\ref{sec:conclusion}.

\section{Banks-Casher relation in a finite volume}
\label{sec:ChPT}

Eigenvalues of the Dirac operator in 
four-dimensional continuum Euclidean space
are pure imaginary (which we denote $i\lambda_1$, $i\lambda_2$, $\cdots$ with real $\lambda_i$'s). 
The spectral density at an eigenvalue $i\lambda$ is defined by 
\begin{eqnarray}
\rho(\lambda) &\equiv& \frac{1}{V}\sum_k^\infty 
\left\langle \delta(\lambda-\lambda_k)\right\rangle,
\end{eqnarray}
where $\langle \cdots \rangle$ denotes an average over the gauge
configuration space. 
Since a nonzero eigenvalue appears as a pair with 
its complex conjugate, $\rho(\lambda)=\rho(-\lambda)$,
we only consider the $\lambda \geq 0$ region in the following.

The chiral condensate $\Sigma$ in the limit of massless quarks and
infinite volume is an order parameter of the chiral symmetry breaking.
Through the Banks-Casher relation \cite{Banks:1979yr}, $\Sigma$ 
is related to $\rho(\lambda)$ as
\begin{equation}
  \lim_{m\to 0}\lim_{V\to \infty}\rho(0) 
  =\frac{\Sigma}{\pi},
\end{equation}
with which one can identify the spontaneous breaking of
chiral symmetry by measuring $\rho(0)$ instead of $\Sigma$.

Even when the volume $V$, 
sea quark masses $\{m_{sea}\}=\{m_u,m_d,m_s,\cdots\}$ and 
$\lambda$ are all finite,
a similar nonperturbative relation
\begin{eqnarray}
  \label{eq:BCgeneral}
\rho(\lambda) &=& - \left.{\rm Re}
\frac{\langle \bar{q}q\rangle}{\pi}\right|_{m_v=i\lambda},
\end{eqnarray}
holds.
Here, $(\cdots)|_{m_v=i\lambda}$ means that the quantity is evaluated
with the valence quark mass analytically continued to a pure imaginary
value $i\lambda$. 
In this relation, the ultraviolet divergence in the definition of the
$\bar{q}q$ operator cancels by taking its real part
at an imaginary value $m_v=i\lambda$
(where the divergent part is pure imaginary),
which is natural because the left hand side of the equation only
refers to low-lying modes and is insensitive to the ultraviolet region
of the dynamics.

We note that the relation (\ref{eq:BCgeneral}) is valid even when the
ensemble average $\langle\cdots\rangle$ is restricted to a given
topological sector of $Q$ in the gauge field configurations 
\cite{Fukaya:2006vs}, that we denote $\langle\cdots\rangle_Q$.
Namely, if we define the spectral density at a given topological sector as
$\rho_Q(\lambda)\equiv(1/V)\sum_k
\left\langle \delta(\lambda-\lambda_k)\right\rangle_Q$,
it is obtained by computing $-{\rm Re}\langle\bar{q}q\rangle_Q/\pi$. 

Although the chiral condensate $\Sigma$ is different from 
$-{\rm Re}\langle \bar{q}q\rangle|_{m_v=i\lambda}$ at finite volumes,
the difference can be described by the low-energy effective theory, 
provided that the energy scales of the theory 
are well below the QCD scale: 
\begin{eqnarray}
\lambda,\;\;\; m_i,\;\;\; 1/V^{1/4} &\ll& \Lambda_{\rm QCD}. 
\end{eqnarray}
It is, therefore, possible to directly compare the lattice QCD
calculation of $\rho(\lambda)$ at finite $\lambda$, $m_i$, $Q$ and $V$ with 
the prediction of the effective theory.
By taking the limit of $m_i\to 0$ after $V\to \infty$ according
to the effective theory and summing over $Q$, one can
reproduce the physical $\rho(\lambda)$, which in the limit of
$\lambda\to 0$ gives $\Sigma$.

In this direction, studies in both lattice QCD 
\cite{DeGrand:2006nv, Lang:2006ab, Hasenfratz:2007yj, 
DeGrand:2007tm, Bar:2010zj, Jansen:2009tt, Hasenfratz:2008ce}
and the low-energy effective theory have been done.
Smilga and Stern \cite{Smilga:1993in} 
and Osborn {\it et al.} \cite{Osborn:1998qb}
calculated the Dirac eigenvalue spectrum in the conventional $p$
expansion of (partially quenched) ChPT to NLO. 
In the vicinity of $\lambda =0$, which corresponds to the limit of
zero valence pion mass, a special treatment of the zero-momentum modes
is needed because the correlation length exceeds the size of the volume.
This special treatment is known as the $\epsilon$ expansion of ChPT,
in which the zero-momentum modes are nonperturbatively integrated
over the $SU(N_f)$ (or $U(N_f)$ when the topological charge $Q$ is fixed)
manifold. 
At the leading order (LO), the finite size effect around $\lambda \sim
0$ was calculated in 
\cite{Damgaard:1997ye,Wilke:1997gf,Akemann:1998ta,Damgaard:2000ah}.
Their results are expressed using the Bessel functions, which has a
$\sim 1/\Sigma V$ gap from zero, reflecting the fact that no
spontaneous symmetry breaking occurs at finite volumes. 

Recently, an interpolation between the $p$  and $\epsilon$ regimes was
considered in \cite{Damgaard:2008zs}. 
The recipe for the calculation is to keep the same counting rule
as in the $p$ expansion but to integrate the zero-modes nonperturbatively
like in the $\epsilon$ expansion.
Partial quenching is performed with the so-called replica trick
so that results at arbitrary nondegenerate set of quark masses can be compared
to lattice QCD.
Using this hybrid method, the results mentioned above
(in the $p$ expansion \cite{Smilga:1993in,Osborn:1998qb} and 
in the $\epsilon$ expansion
\cite{Damgaard:1997ye,Wilke:1997gf,Akemann:1998ta,Damgaard:2000ah})
are smoothly connected.
Comparison with the lattice data is, therefore, no longer limited in
either the $\epsilon$  or $p$ regimes, and more precise determination of
$\Sigma$ is possible.

Here we briefly reproduce the result of \cite{Damgaard:2008zs}
where we consider a general theory with $N_f$ flavors of sea quarks.
The spectral density in a fixed topological sector of $Q$ is given by
\begin{equation}
  \label{eq:rho}
  \rho_Q(\lambda) = 
  \Sigma_{\rm eff}
  \hat{\rho}^\epsilon_Q(\lambda \Sigma_{\rm eff}V,\{m_{sea}\Sigma_{\rm eff}V\})
  + \rho^p(\lambda,\{m_{sea}\}),
\end{equation}
where two terms
$\hat{\rho}^\epsilon_Q(\lambda \Sigma_{\rm eff}V,\{m_{sea}\Sigma_{\rm eff}V\})$
and $\rho^p(\lambda,\{m_{sea}\})$ are given in the following.
$\Sigma_{\rm eff}$ includes the leading finite quark mass correction
to $\Sigma$ that modifies the overall normalization of the
spectrum, and is therefore called the {\it effective} chiral condensate.

The spectrum of the near-zero modes ($\lambda\sim 1/\Sigma V$)
is mainly affected by the zero-momentum pion fields.
The first term in (\ref{eq:rho}) has the same functional form as the
one at the leading order of the $\epsilon$ expansion
\cite{Damgaard:1997ye,Wilke:1997gf,Akemann:1998ta,Damgaard:2000ah},
which is expressed in terms of dimensionless combinations
$\lambda\Sigma_{\rm eff} V$ and 
$\{m_{sea}\Sigma_{\rm eff}V\}=\{m_1\Sigma_{\rm eff}V, \cdots,$
$m_{N_f}\Sigma_{\rm eff}V\}$:
\begin{eqnarray}
\hat{\rho}^\epsilon_Q(\zeta,\{\mu_{sea}\})
  \equiv C_2\frac{|\zeta|}{2\prod^{N_f}_f(\zeta^2 + \mu^2_f)}
  \frac{\det\tilde{\mathcal{B}}}{\det\mathcal{A}},
\end{eqnarray}
with $N_f\times N_f$ matrix $\mathcal{A}$ and 
$(N_f+2)\times(N_f+2)$ matrix $\tilde{\mathcal{B}}$ defined by
$\mathcal{A}_{ij}= \mu_i^{j-1}I_{Q+j-1}(\mu_i)$ and 
$\tilde{\mathcal{B}}_{1j} =  \zeta^{j-2}J_{Q+j-2}(\zeta)$,
$\tilde{\mathcal{B}}_{2j} =  \zeta^{j-1}J_{Q+j-1}(\zeta)$,
$\tilde{\mathcal{B}}_{ij} = (-\mu_{i-2})^{j-1}I_{Q+j-1}(\mu_{i-2})$ 
$(i\neq 1,2)$, respectively
($J_k$'s and $I_{l}$'s denote the (modified) Bessel functions.).
The phase factor $C_2$ is 1 for $N_f=2$ and 3.

The second term in (\ref{eq:rho}) is a logarithmic 
NLO correction (chiral-logarithms) which is also partly seen 
in the conventional $p$ expansion \cite{Osborn:1998qb}.
With the meson mass $M_{ij}^2\equiv (m_i+m_j)\Sigma/F^2$, which is 
made of either sea quark ($f$) or valence quark ($v$), it is
given by\footnote{In this paper, we use simplified notations: 
$\bar{\Delta}(M^2)$ and $\bar{G}(M^2)$ correspond to $\bar{\Delta}(0,M^2)$ and
$\bar{G}(0,M^2,M^2)$ in \cite{Damgaard:2008zs}, respectively.}
\begin{eqnarray}
  \label{eq:rhop}
  \rho^p(\lambda,\{m_{sea}\}) &\equiv&
  -\frac{\Sigma}{\pi F^2}{\rm Re}\left.\left[
  \sum^{N_f}_f (\bar{\Delta}(M^2_{fv})
  -\bar{\Delta}(M^2_{ff}/2))
  -(\bar{G}(M^2_{vv})-\bar{G}(0))
  \right]\right|_{m_v=i\lambda}.
\end{eqnarray}
where 
\begin{eqnarray}
\bar{G}(M^2) &=& \left\{
\begin{array}{l}
\displaystyle\frac{1}{2}\left[ 
\bar{\Delta}(M^2) +
(M^2-M_{ud}^2)\partial_{M^2}\bar{\Delta}(M^2)\right] 
\hspace{1in}(N_f=2),\\\\
\displaystyle\frac{1}{3}\left[ 
-\frac{2(M^2_{ud}-M^2_{ss})^2}{9(M^2-M^2_{\eta})^2}\bar{\Delta}(M_\eta^2)
+\left(1 + \frac{2(M_{ud}^2-M_{ss}^2)^2}{9(M^2-M^2_{\eta})^2}\right)
\bar{\Delta}(M^2)
\right.
\\\hspace{1in}\left.
\displaystyle+\frac{(M^2-M^2_{ud})(M^2-M^2_{ss})}{(M^2-M^2_{\eta})}
\partial_{M^2}\bar{\Delta}(M^2)\right] 
\hspace{0.15in}(N_f=2+1,\;3),\\
\end{array}\right.
\\\bar{\Delta}(M^2)&=& 
\frac{M^2}{16\pi^2}\ln \frac{M^2}{\mu_{sub}^2}
+\bar{g}_1(M^2).
\label{eq:Deltabar}
\end{eqnarray}
Here, the physical meson masses are given by the leading
order relations
$M_{ud}^2=2m_u\Sigma/F^2=2m_d\Sigma/F^2$, $M_{ss}^2=2m_s\Sigma/F^2$
and $M_\eta^2=(M_{ud}^2+2M_{ss}^2)/3$.
The scale $\mu_{sub}$ (= 770~MeV in this work) is a subtraction scale.
The function given by $\bar{g}_1(M^2)=g_1(M^2)-1/M^2V$ denotes
a finite volume correction from nonzero momentum pion modes.
In the $p$ expansion, it is expressed by the modified Bessel function $K_1$
\cite{Bernard:2001yj}
while in the $\epsilon$ expansion a polynomial expression is used
\cite{Hasenfratz:1989pk}.
In this study, we need the both expressions:
\begin{eqnarray}
\label{eq:g1nume}
\bar{g}_1(M^2) = \left\{
\begin{array}{lc}
\displaystyle\sum_{a
\neq 0,\;|n_i| \leq n^{max}_1} 
\frac{\sqrt{M^2}}{4\pi^2|a|}K_1(\sqrt{M^2}|a|)-\frac{1}{M^2V}& (|M|L>2)\\\\
\displaystyle-\frac{M^2}{16\pi^2}\ln (M^2 V^{1/2})
-\sum^{n^{max}_2}_{n=1}\frac{\beta_n}{(n-1)!} M^{2(n-1)}V^{(n-2)/2} &(|M|L\leq 2)
\end{array}
\right. ,
\end{eqnarray}
where $a_\mu$ denotes a four-vector $a_\mu=(n_1L,n_2L,n_3L,n_4T)$ with
integer $n_i$'s and $\beta_i$'s are the {\it shape coefficients} defined 
in \cite{Hasenfratz:1989pk}.
Their formula and numerical values for the first several $\beta_n$'s
are summarized in Appendix~\ref{app:beta}.
In our numerical study, we truncate the sum at $n_1^{max}=7$ and
$n_2^{max}=300$,  
which indeed shows a good convergence around the threshold $|M|L= 2$.
We note that both  $\bar{\Delta}(M^2)$ and $\bar{G}(M^2)$ are
finite even in the limit of $M\to 0$. 

The effective chiral condensate $\Sigma_{\rm eff}$ in (\ref{eq:rho}) is given by
\begin{equation}
  \label{eq:Sigmaeff}
  \Sigma_{\rm eff} = 
  \Sigma \left[1-\frac{1}{F^2}\left(
      \sum^{N_f}_f \bar{\Delta}(M^2_{ff}/2)-\bar{G}(0)
      -16L^r_6\sum^{N_f}_f M^2_{ff}\right)\right],
\end{equation}
where $L_6^r$ (renormalized at $\mu_{sub}$)
is one of the low-energy constants at NLO 
\cite{Gasser:1983yg}.
From the sea quark mass dependence of $\Sigma_{\rm eff}$,
one can determine $\Sigma$ as well as $F$ and $L_6^r$.

In the expression (\ref{eq:rho}), dependence on the topological
charge $Q$ is encoded only in the first term
$\Sigma_{\rm eff}\hat{\rho}^\epsilon_Q(\lambda\Sigma_{\rm eff}V,
\{m_{sea}\Sigma_{\rm eff}V\})$ and 
the second term $\rho^p(\lambda,\{m_{sea}\})$ does not depend on $Q$
since it is a contribution from nonzero momentum modes.
On the other hand, the chiral logarithm manifests itself in 
the both terms
through $\bar{\Delta}(M^2)$.
Since $M^2$ could also contain $\lambda$ through $m_v=i\lambda$, 
the spectral density shows a nonanalytic functional form.

The above ChPT results are subject to higher order corrections in the
$p$ expansion, for which the expansion parameter is either $M^2/F^2$
or $1/(FV^{1/4})^2$.
Although the zero-mode contribution is treated nonperturbatively, 
there are two-loop contribution of nonzero momentum modes that could also couple to
the zero-mode and introduce different types of group integrals.
At the two-loop level, these contributions may have the order
$M^4/F^4$, $M^2/(F^4V^{1/2})$ or $1/(F^4V)$, whose coefficients are
unknown. 
We therefore need to carefully check the convergence of the 
expansion at NLO for our parameter sets.
In the following analysis, we test the NLO formula with various
sets of quark masses, as well as different 
lattice volumes ($L$ = 16, 24) for the $N_f=2+1$ runs and
different topological sectors for the $N_f=2$ runs, 
in order to confirm the convergence.

\section{Lattice QCD simulations}
\label{sec:lattice}

Numerical simulations of lattice QCD are performed 
with the Iwasaki gauge action \cite{Iwasaki:1985we} at $\beta=2.3$ 
(except for the run of $N_f=2$ QCD at $m_{ud}=0.002$ 
for which we choose $\beta=2.35$)
including 2, 2+1 ($m_s$ fixed), and 3 ($m_{ud}=m_s$) flavors of
dynamical quarks. 
For the quark action, we employ the overlap-Dirac operator
\cite{Neuberger:1997fp} 
\begin{equation}
  \label{eq:ov}
  D(m) = 
  \left(m_0+\frac{m}{2}\right)+
  \left(m_0-\frac{m}{2}\right)
  \gamma_5 \mbox{sgn}[H_W(-m_0)],
\end{equation}
where $m$ denotes the quark mass and
$H_W\equiv\gamma_5D_W(-m_0)$ is the Hermitian
Wilson-Dirac operator with a large negative mass $-m_0$.
We take $m_0=1.6$ throughout our simulations
(here and in the following the parameters are 
given in the lattice unit.).
For the details of numerical implementation 
of the overlap-Dirac operator,
we refer to our previous paper \cite{Aoki:2008tq}.

It is known that the numerical cost for the dynamical 
simulation of the overlap fermions becomes prohibitively large
when $H_W(-m_0)$ has (near) zero-modes. 
To avoid this problem, we introduce extra Wilson fermions and
associated twisted mass bosonic spinors to generate a weight
\begin{equation}
  \label{eq:detHw}
  \frac{\det[H_W^2(-m_0)]}{\det[H_W^2(-m_0)+m_t^2]},
\end{equation}
in the functional integrals
\cite{Izubuchi:2002pq,Vranas:2006zk,Fukaya:2006vs}.
Both of these fermions and ghosts are unphysical 
as their masses are of order of the lattice
cutoff, and do not affect low-energy physics.
The numerator suppresses the appearance of near-zero modes, 
while the denominator cancels unwanted effects from high modes.
The {\it twisted-mass} parameter $m_t$ controls the value of
threshold below which the eigenmodes are suppressed.
In our numerical studies, we set $m_t$ = 0.2.

With the determinant (\ref{eq:detHw}) 
the index of the overlap-Dirac operator,
or the topological charge in the continuum limit \cite{Hasenfratz:1998ri}, 
never changes from its initial value
during the molecular dynamics steps 
since its change always requires crossing
zero eigenvalue of $H_W(-m_0)$.
In this work the simulations are mainly performed in the trivial
topological sector, $Q=0$.
In order to check the topological charge dependence, we also 
carry out independent simulations at $Q=+1$, $-2$ and $-4$
at several sets of parameters.

\begin{table}[htb]  
  \centering
  \begin{tabular}{cccccccccc}
    \hline\hline
$N_f$ & $V$\hspace{0.2in} & $\beta$ & $a^{-1}$(GeV) &  $m_{ud}$ & $m_s$ & $Q$ & $N_{trj}$ & $\tau_{trj}$ & $N_{auto}$\\
    \hline
2 &  $16^3\times32$\hspace{0.2in} & 2.35 & 1.776(38) & 0.002 & $\infty$ & 0 & 4680  & 0.5 & 34(12) \\
\cline{3-10}
  &                 & 2.30 & 1.667(17) & 0.015 & $\infty$ & 0 & 10000 & 0.5 & 48(21)\\
  &                 &      &           & 0.025 & $\infty$ & 0 & 10000 & 0.5 & 38(16)\\
  &                 &      &           & 0.035 & $\infty$ & 0 & 10000 & 0.5 & 28(12)\\
  &                 &      &           & 0.050 & $\infty$ & 0 & 10000 & 0.5 & 24(9)\\
  &                 &      &           & 0.050 & $\infty$ &-2 & 5000  & 0.5 & 50(24)\\
  &                 &      &           & 0.050 & $\infty$ &-4 & 5000  & 0.5 & 34(16)\\
  &                 &      &           & 0.070 & $\infty$ & 0 & 10000 & 0.5 & 23(10)\\
  &                 &      &           & 0.100 & $\infty$ & 0 & 10000 & 0.5 & 9(3)\\
\hline
2+1 &$16^3\times48$\hspace{0.2in} & 2.30 & 1.759(10) & 0.002 & 0.080 & 0 & 5000 & 0.5 & 17(9) \\
  &                &      &           & 0.015 & 0.080 & 0 & 2500 & 1.0 & 15(6) \\
  &                &      &           & 0.015 & 0.080 & 1 & 1800 & 1.0 & 5(1) \\
  &                &      &           & 0.025 & 0.080 & 0 & 2500 & 1.0 & 11(5)\\
  &                &      &           & 0.035 & 0.080 & 0 & 2500 & 1.0 & 24(11)\\
  &                &      &           & 0.050 & 0.080 & 0 & 2500 & 1.0 & 9(5)\\
\cline{5-10}
  &                &      &           & 0.015 & 0.100 & 0 & 2500 & 1.0 & 5(3)\\
  &                &      &           & 0.025 & 0.100 & 0 & 2500 & 1.0 & 10(3)\\
  &                &      &           & 0.035 & 0.100 & 0 & 2500 & 1.0 & 34(21)\\
  &                &      &           & 0.050 & 0.100 & 0 & 2500 & 1.0 & 5(3) \\
\cline{2-10}
  & $24^3\times48$\hspace{0.2in} & 2.30 & 1.759(10) & 0.015 & 0.080 & 0 & 2500 & 1.0 & 2(1)\\
  &                &      &           & 0.025 & 0.080 & 0 & 2500 & 1.0 & 3(2) \\
    \hline
3 & $16^3\times48$\hspace{0.2in} & 2.30 & 1.759(10) & 0.025 & 0.025 & 0 & 2500 & 1.0 & 4(1)\\
  &                &      &           & 0.035 & 0.035 & 0 & 2500 & 1.0 & 5(1) \\
  &                &      &           & 0.080 & 0.080 & 0 & 2500 & 1.0 & 20(12)\\
  &                &      &           & 0.100 & 0.100 & 0 & 2500 & 1.0 & 24(15)\\
\hline
  \end{tabular}
  \caption{
    Summary of lattice parameters.
    For the 2-, 2+1- and 3-flavor runs, the values of $\beta$,
    $a^{-1}$, $(m_{ud},m_s)$ and $Q$ are listed.
    Other parameters are those for the HMC simulations:
    $N_{trj}$ denotes the number of trajectory,
    $\tau_{trj}$ is the unit trajectory length, and 
    $N_{auto}$ denotes the integrated auto-correlation length (number
    of trajectories) of the lowest eigenvalue.
  }
  \label{tab:simprm}
\end{table}

Simulation parameters are summarized in Table~\ref{tab:simprm}.
The lattice size is $V=16^3\times 32$ for the $N_f=2$ runs, 
while it is $V=16^3\times 48$ for the main $N_f=2+1$ and $N_f=3$ runs.
In order to investigate the finite volume scaling, we also simulate on a
$24^3 \times 48$ lattice at the same lattice spacing
with two choices of sea quark masses $(m_{ud},m_s)=(0.015, 0.080)$
and  $(0.025, 0.080)$.
For the determination of the lattice spacing $a$ = 0.11--0.12~fm ($a^{-1}=$ 1.7 -- 1.8 GeV),
we choose the $\Omega$-baryon mass as the input for the $N_f=2+1$ and 3 ensembles \cite{JLQCD:prep},
while it is determined from the heavy
quark potential with an input $r_0$ = 0.49~fm for the $N_f=2$ case.
Our lattice size is then estimated as $L\sim 1.8$~fm for the $N_f=2+1$ runs, 
and $L\sim 1.9$~fm for the $N_f=2$ runs.

In the $N_f=2$ runs, seven different values of the 
up and down quark mass $m_{ud}$ are taken.
For the $N_f=2+1$ runs, 
we choose two different values of strange quark mass
$m_s$ (= 0.080 and 0.100) and 
six (for the former) or five (for the latter) values of
$m_{ud}$ are chosen. 
For the degenerate $N_f=3$ flavor runs, we take
four different values of $m_{ud}=m_s$.
Note that the lightest up and down quark mass $m_{ud}=0.002$
in the $N_f=2$ or $N_f=2+1$ runs roughly corresponds to 3~MeV in the
physical unit, with which pions are in the $\epsilon$ regime while
kaons still remain in the $p$ regime.

We compute 50--80 pairs of low-lying eigenvalues 
(that we denote $\lambda^{ov}$'s) 
of the massless overlap-Dirac operator $D(0)$
at every 5 or 10 (depending on the parameters) trajectories.
In the calculation of the eigenvalues, we employ the implicitly
restarted Lanczos algorithm for the chirally projected operator 
$P_+\,D(0)\,P_+$ (where $P_+\!=\!(1+\gamma_5)/2$) of 
which eigenvalue corresponds to ${\rm Re} \lambda^{ov}$.
From each eigenvalue of $P_+\,D(0)\,P_+$,  
the eigenvalue $\lambda^{ov}$, as well as its complex conjugate, 
are extracted through the relation
$|1-\lambda^{ov}/m_0|^2=1$. 
In order to compare with ChPT, every complex eigenvalue
$\lambda^{ov}$ is mapped onto the imaginary axis as
$\lambda\equiv\mathrm{Im}\lambda^{ov}/(1-\mathrm{Re}\lambda^{ov}/(2m_0))$.
The difference between $\lambda$ and $\mbox{Im}\lambda^{ov}$ is a
discretization effect, which is negligible (within 1\%) for $|\lambda^{ov}| <$~0.03.
In the analysis, we consider positive $\lambda$ only.

For each run, 1,800--10,000 (depending on the parameters) trajectories
are accumulated  using the hybrid Monte Carlo algorithm.
The integrated auto-correlation time $N_{auto}$ of the lowest $\lambda$
in the unit of
the trajectory length $\tau_{trj}$ is also
listed in Table~\ref{tab:simprm}. 
Because of its infra-red nature, the lowest Dirac eigenvalue is
expected to be most difficult to decorrelate and thus has the longest
auto-correlation time.
The measurement is not stable and the statistical error is as large as
50\% in some cases, but $N_{auto}$ is typically $O(50)$ or less.
In the following analysis, the statistical error for the spectral
density and other quantities is estimated by the jackknife
method after binning the data in every 100 trajectories. 

Details of configuration generation and other quantities 
will be reported in a separate paper.

\section{Extraction of LECs at each set of
  sea quark masses} 
\label{sec:extraction}

Although a global fit of the lattice data for spectral density to the
NLO ChPT formula is possible in principle, we prefer to simplify the analysis, 
for better understanding of numerical sensitivity of the lattice
data and the errors in the final result.
We first consider the mode number below $\lambda$, 
or the integrated eigenvalue density,
\begin{eqnarray}
  \label{eq:cum}
  N_Q(\lambda) &\equiv& V\int^{\lambda}_0 d\lambda^\prime \rho_Q (\lambda^\prime),
\end{eqnarray}
at each set of sea quark masses.
The analytic ChPT result (\ref{eq:rho}) is also integrated numerically
from 0 to $\lambda$.
In the second term of (\ref{eq:rho}) we can replace 
$\Sigma/F^2$ by $\Sigma_{\rm eff}/F^2$ as their difference is a higher
order effect.
Then, there are two unknown parameters in the formula:
$\Sigma_{\rm eff}$ and $F$.

The data points of $N_Q(\lambda)$ at two reference values
of $\lambda$ are hence sufficient to determine the parameters.
As the reference points we take $\lambda$ = 0.004 ($\sim$ 7~MeV) and 
0.017 ($\sim$ 30~MeV) except for the case with $m_{ud}$ = 0.002, for
which we choose $\lambda$ = 0.0125 and 0.017 ($N_f=2$) or
0.010 and 0.017 ($N_f=2+1$).
For the $Q\neq 0$ runs we take $\lambda$ = 0.01 and 0.02. 
Effectively, the lower $\lambda$ point determines $\Sigma_{\rm eff}$,
while the other point is more sensitive to the NLO effects that
contain $1/F^2$.
We check that the resulting values of $\Sigma_{\rm eff}$ and $F$ are
stable against the change of the reference points by varying them 
by a factor of 2 or 3 while keeping the higher point less than 0.030
to avoid possible higher order corrections.
The reference points are different for the $\epsilon$ regime runs and
for the $Q\neq 0$ runs, because the small eigenvalues are highly
suppressed (the lowest eigenvalue is larger than 0.004) for these cases.

For the $N_f$ = 2+1 lattice data, we test both the $N_f=2+1$ and $N_f=2$
ChPT formulas. 
For the latter case, the strange quark is assumed to be decoupled from
the theory, which we call {\it reduced} $N_f=2$ ChPT. 

Numerical results are listed in Table~\ref{tab:SigmaeffFeff}.
Before moving to further analysis of the results
let us describe our observations for the spectral function.

\begin{figure}[tbp]
  \centering
  \includegraphics[width=13cm]{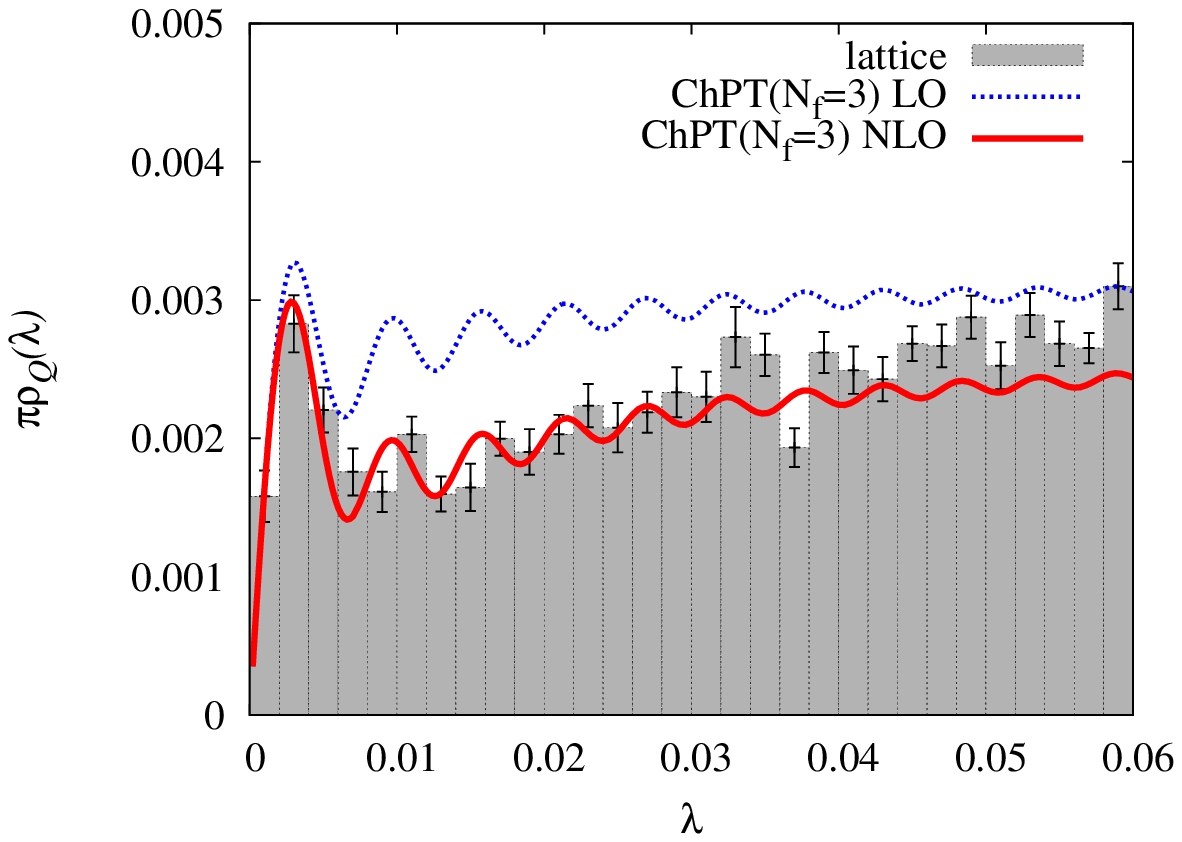}
  \includegraphics[width=13cm]{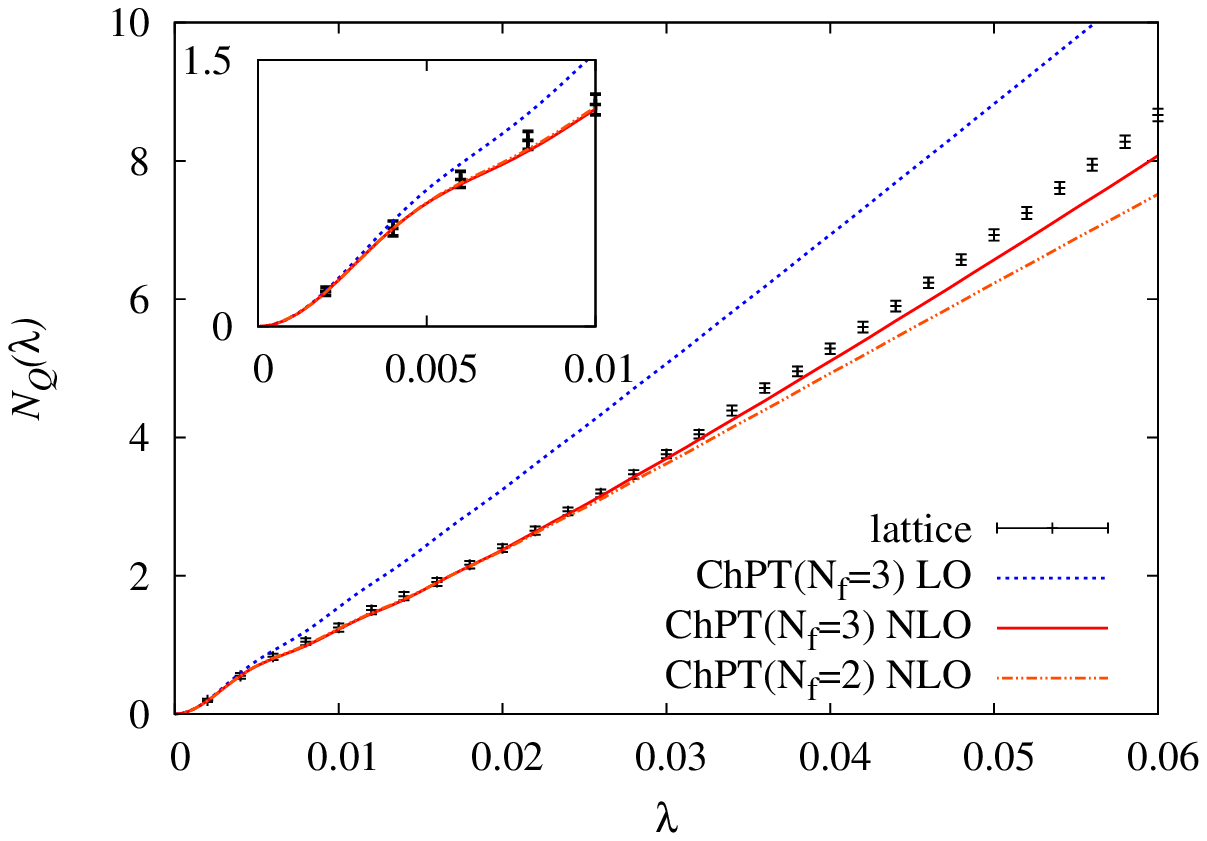}
  \caption{
    $N_f=$2+1 lattice QCD results 
    for the spectral density $\rho_Q(\lambda)$ (top panel) and the
    mode number $N_Q(\lambda)$ (bottom panel) of the Dirac operator at
    $m_{ud}$ = 0.015, $m_s$ = 0.080 and $Q=0$. 
    The lattice data (histogram (top) or solid symbols (bottom)) are
    compared with the NLO ChPT formula drawn by solid curves.
    For comparison, the prediction of the leading-order
    $\epsilon$ expansion (dotted curves) and that of 
    the NLO formula but with $N_f=2$ flavors (dashed) are also shown. 
  }
  \label{fig:rho015}
\end{figure}

\begin{figure}[tbp]
  \centering
  \includegraphics[width=13cm]{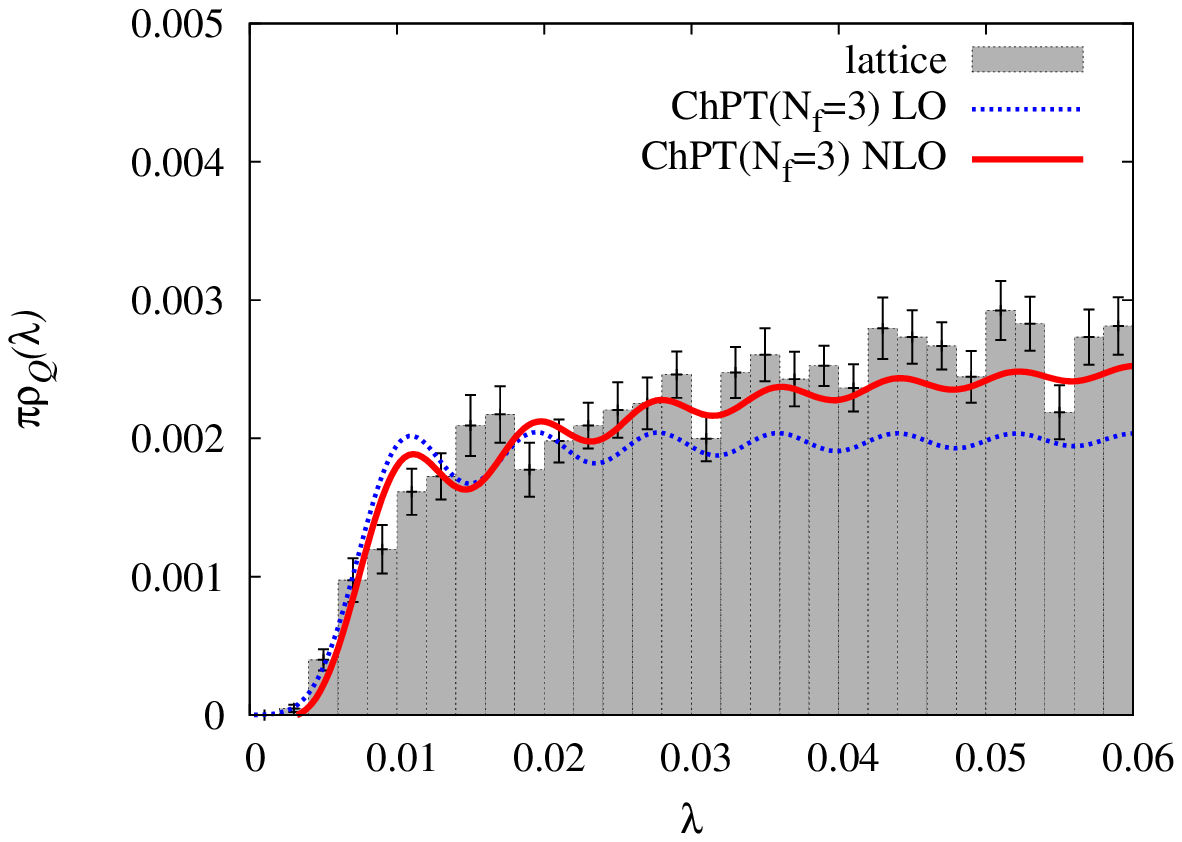}
  \includegraphics[width=13cm]{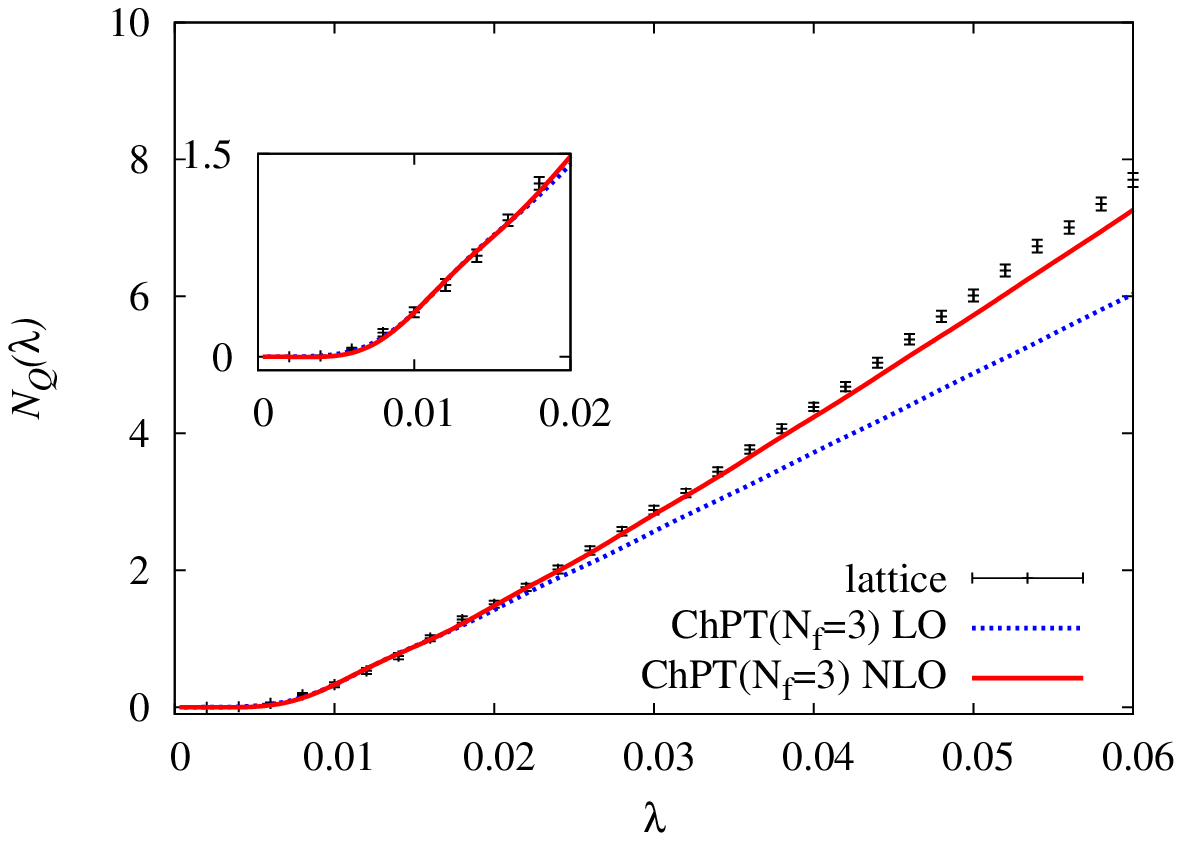}
  \caption{
    Same as Figure~\protect\ref{fig:rho015}, but at
    $m_{ud}=0.002$ and $m_s=0.080$.
  }
  \label{fig:rho002}
\end{figure}

Figures~\ref{fig:rho015}--\ref{fig:V}
show the spectral density and its integral (\ref{eq:cum}) obtained in
our lattice simulations. 
A typical example is that for $N_f=2+1$ QCD in the $p$ regime
(Figure~\ref{fig:rho015}).
The upper panel shows the histogram plot of the spectral function
$\rho_Q(\lambda)$ as a function of $\lambda$.
The lattice data for each bin have a jackknife estimate of the
statistical error.
The solid (red) curve represents the NLO ChPT formula,
while the dotted (blue) curve corresponds to the leading order result
(in the $\epsilon$ expansion).
Since the first reference point is 0.004, it probes the first peak,
which corresponds to the lowest eigenvalue.
Starting from the second peak, the effect of the NLO term appears,
as clearly seen from the difference of the two curves.
Therefore, by taking the second reference point at 0.017, the data
have enough sensitivity to the NLO parameter $1/F^2$.
This observation is of course specific to the particular volume of our
lattice; on larger lattices, the peaks move toward the origin and the
impact of the NLO term would become less significant on the second or
third peaks (see Figure~\ref{fig:V}).

The agreement with the formula can be seen more clearly by looking at
$N_Q(\lambda)$, the mode number below $\lambda$ (lower panel).
The lattice data depart from the leading-order curve (dotted),
which corresponds to the first term 
$\Sigma_{\rm eff}\hat{\rho}_Q^\epsilon$ in (\ref{eq:rho}), 
at around 0.005 (see the inset).
Then, the data follow the nontrivial functional form of
the NLO formula (solid), which comes from the chiral logarithm originating
from pion loops.
The NLO formula works precisely up to $\lambda\sim 0.025$,
and the deviation is still within two sigma at $\lambda\sim 0.04$,
which is about the half of the physical strange quark mass $m^{\rm phys}_s$.
This is a typical range where the NLO ChPT is valid, and the higher
order corrections would become sizable above this value.
In contrast to a recent work by Giusti and L\"uscher
\cite{Giusti:2008vb}, where they take a wider range of $\lambda$ (up
to $\lambda\sim$ 95~MeV) into the analysis, we conservatively choose
the reference points below 30~MeV, so that (partially quenched) ChPT with
an imaginary valence quark mass $i\lambda$ can be safely applied.

In Figure~\ref{fig:rho015}, the difference between the $N_f=3$ and
$N_f=2$ (dashed curve) formulas
is not sizable below $\lambda\sim$ 0.03--0.04.
This is natural because the strange quark (with $m_s=0.08$) decouples
from the dynamics of the low-lying modes.
As a result, the extraction of $\Sigma_{\rm eff}$ does not
significantly depend on the formula we use ($N_f=3$ or $N_f=2$).

The convergence of the chiral expansion is better in the
$\epsilon$ regime as shown in Figure~\ref{fig:rho002}, in which the
$N_f=2+1$ lattice data at $m_{ud}$ = 0.002 and $m_s$ = 0.080 are plotted.
In the plot of the mode number (lower panel), the LO and NLO curves
coincide up to $\lambda\sim$ 0.025. 
Beyond this value, we observe some deviation, which is also seen in
the histogram plot (upper panel).

The NLO ChPT correction in this work explains 
the disagreement of the lattice data with the expectation
from the random matrix theory found in our previous work
\cite{Fukaya:2007fb,Fukaya:2007yv}. 
Namely, if we adjust the parameter $\Sigma_{\rm eff}$ using the lowest
eigenvalue distribution (the first peak of the histogram), then the
second peak would not agree at the leading order.
Indeed, the NLO contribution is responsible for this.

\begin{table}[tbp]
  \centering
  \begin{tabular}{cccccccc}
    \hline\hline
    \multirow{2}{*}{$N_f$ (lattice)} & 
    \multirow{2}{*}{$m_{ud}$} & 
    \multirow{2}{*}{$m_s$} &
    \multicolumn{2}{c}{$N_f$=3 ChPT} & 
    \multicolumn{2}{c}{(reduced) $N_f$=2 ChPT}& 
    \multirow{2}{*}{comment}\\
    & & &  $\Sigma_{\rm eff}$ & $F$ & $\Sigma_{\rm eff}$ & $F$ & \\
    \hline
2&     0.002 & $\infty$
     & -- &   -- & 0.00218(19) &  0.059(65) & ($\beta$=2.35)\\
&     0.015 & $\infty$ 
     & -- & -- & 0.00362(15) & 0.0527(20)\\
&    0.025 &  $\infty$ 
    & -- & --  & 0.00353(15) & 0.0664(90)\\
&    0.035 & $\infty$ 
    & -- & -- & 0.00382(14) & 0.0681(64)\\
&    0.050 & $\infty$ 
    & -- & -- & 0.00449(15) & 0.0644(20)\\
&    0.050 & $\infty$ 
    & -- & -- & 0.00400(16) & 0.0728(60) & ($Q$=-2)\\   
&    0.050 & $\infty$ 
    & -- & -- & 0.00482(19) & 0.0636(19) & ($Q$=-4)\\   
&     0.070 & $\infty$ 
    & -- & -- & 0.00480(15) & 0.0707(23)\\
&     0.100 & $\infty$ 
    & -- & -- & 0.00478(12) & 0.0862(72)\\
\hline
2+1&     0.002 &0.080 
     & 0.00204(07) &   0.0469(102) & 0.00204(05) &  0.0425(49) \\
&     0.015 & 0.080
     & 0.00314(18) & 0.0536(15)& 0.00305(17) & 0.0551(16)\\
&    0.015 & 0.080 
    &0.00354(48) & 0.0521(25) & 0.00319(58)& 0.0558(62) & ($Q=1$)\\
&    0.015 & 0.080 
    &0.00273(06) & 0.0520(25) & 0.00270(06)& 0.0545(26) & ($L$=24)\\
&    0.025 &0.080
    & 0.00333(18) & 0.0624(20) & 0.00326(18) & 0.0647(20)\\
&    0.025 &0.080 & 0.00299(06) & 0.0600(23) & 0.00297(05) & 0.0629(24) & ($L$=24)\\
&    0.035 & 0.080
    & 0.00404(39) & 0.0636(17) & 0.00393(36) & 0.0666(16)\\
&    0.050 & 0.080 
    & 0.00423(22) & 0.0696(16)& 0.00413(21) & 0.0738(16)\\
&    0.015 &0.100 
    & 0.00309(14) & 0.0564(19)& 0.00303(13) & 0.0578(19)\\ 
&    0.025 &0.100 
    & 0.00349(20) & 0.0622(17)& 0.00342(19) & 0.0642(17)\\ 
&    0.035 &0.100 
    & 0.00418(40) & 0.0647(14)& 0.00409(38) & 0.0673(14)\\ 
&    0.050 &0.100 
    & 0.00383(13) & 0.0713(16)& 0.00376(13) & 0.0747(16)\\ 
    \hline
3 &    0.025 & 0.025 
    & 0.00335(23) & 0.0531(10)& -- & -- \\  
&    0.035 & 0.035 
    & 0.00334(21) & 0.0612(24)& -- & --\\  
&    0.080 & 0.080 
    & 0.00453(23) & 0.0767(14)& -- & --\\    
&    0.100 &0.100 
    & 0.00520(22) & 0.0835(22)& -- & --\\
    \hline
  \end{tabular}
  \caption{
    Extracted values of $\Sigma_{\rm eff}$ and $F$ using 
    $N_f=3$ and $N_f=2$ ChPT.
  }
  \label{tab:SigmaeffFeff}
\end{table}

\section{Scaling tests with various parameters}
\label{sec:scaling}

\subsection{Sea quark masses}

\begin{figure}[tbp]
  \centering
  \includegraphics[width=14cm]{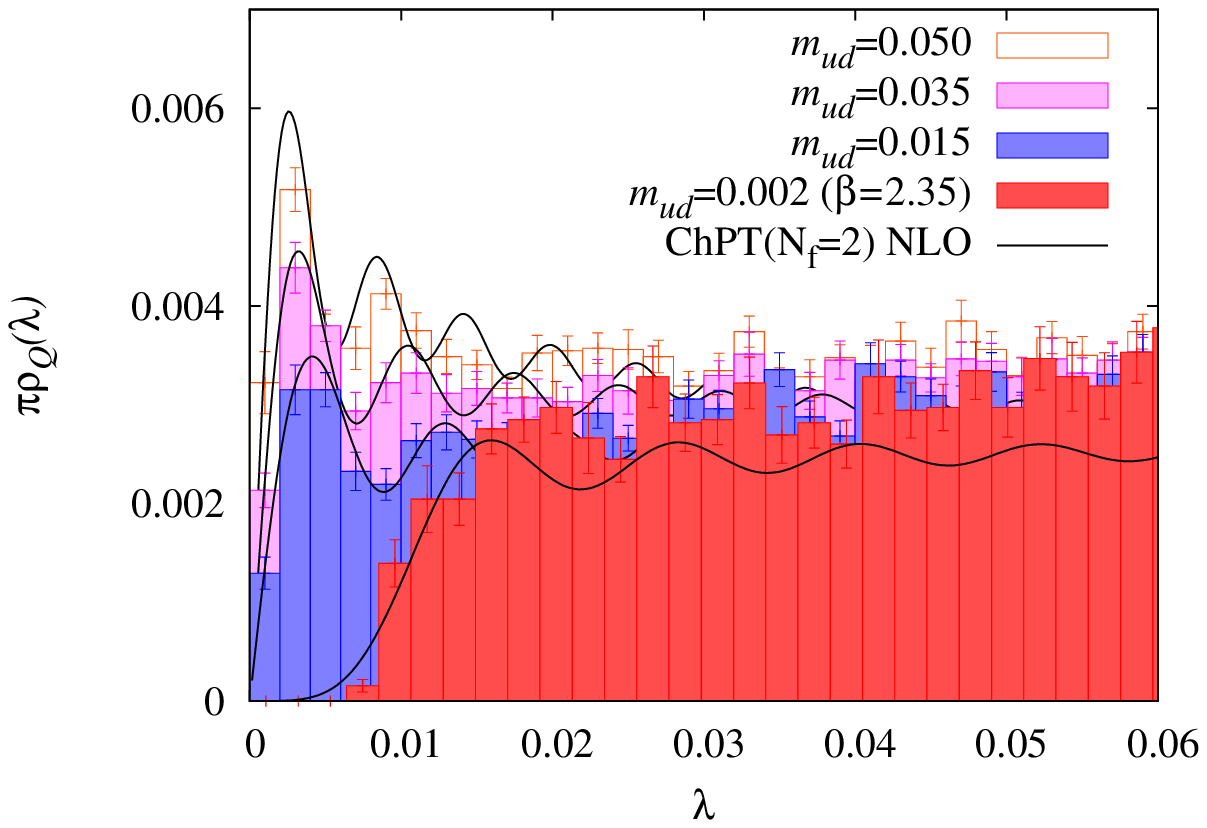}
  \includegraphics[width=14cm]{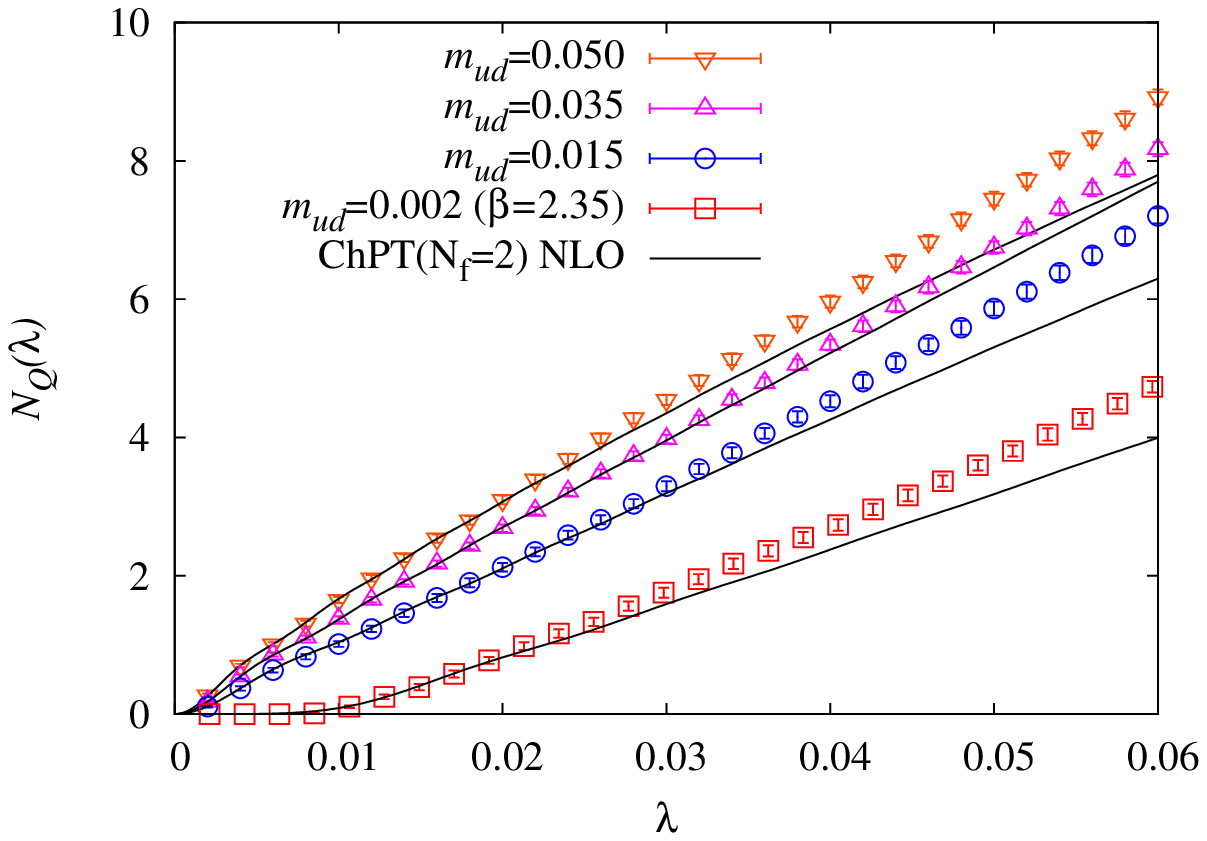}
  \caption{
    $N_f=2$ lattice QCD results 
    for the spectral density $\rho_Q(\lambda)$ (top panel) and the
    mode number $N_Q(\lambda)$ (bottom panel) of the Dirac operator
    at various sea quark masses.
    The global topological charge is fixed to zero.
    The NLO ChPT ($N_f=2$) results are drawn by solid curves.
    Note that the $\beta=2.35$ data are rescaled as
    $\lambda \to 1.065 \lambda$ and $\rho \to 1.209 \rho$ 
    according to the difference of the lattice spacing $a$.
  }
  \label{fig:Nf2}
\end{figure}

\begin{figure}[tbp]
  \centering
  \includegraphics[width=14cm]{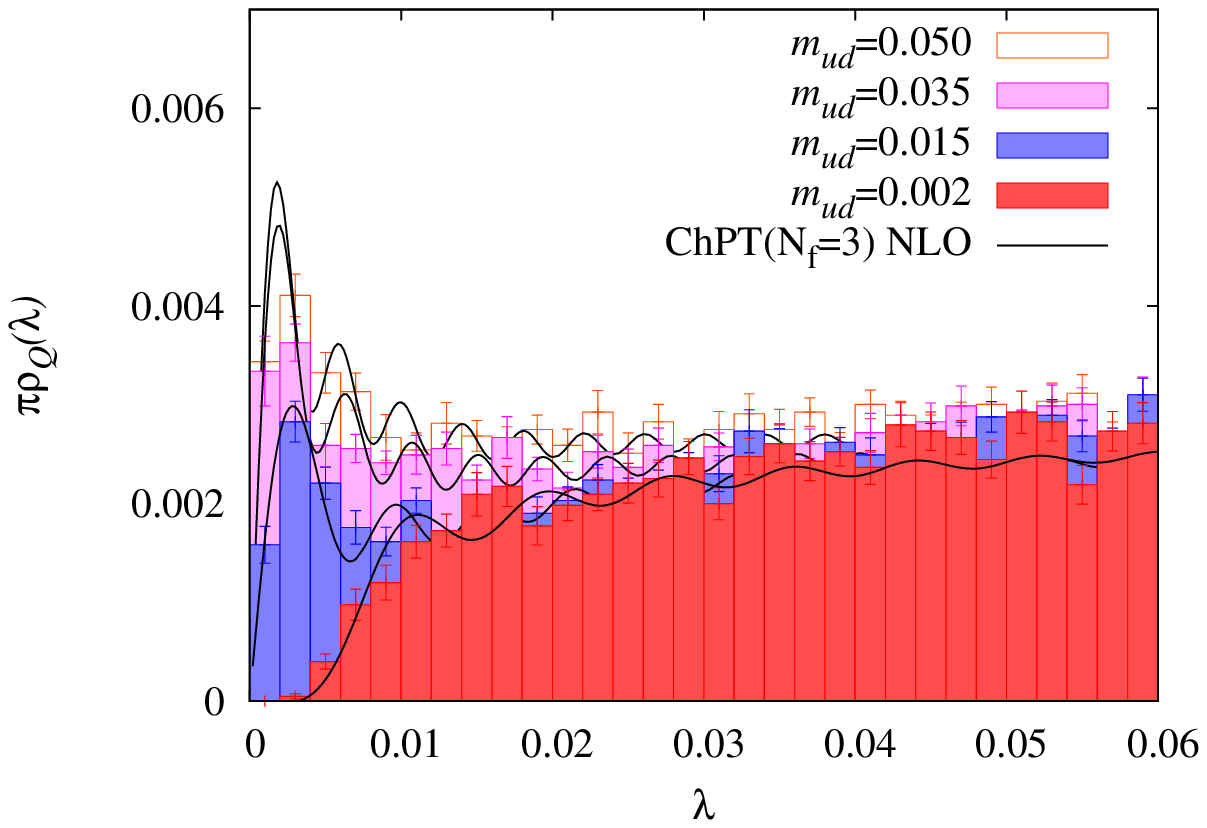}
  \includegraphics[width=14cm]{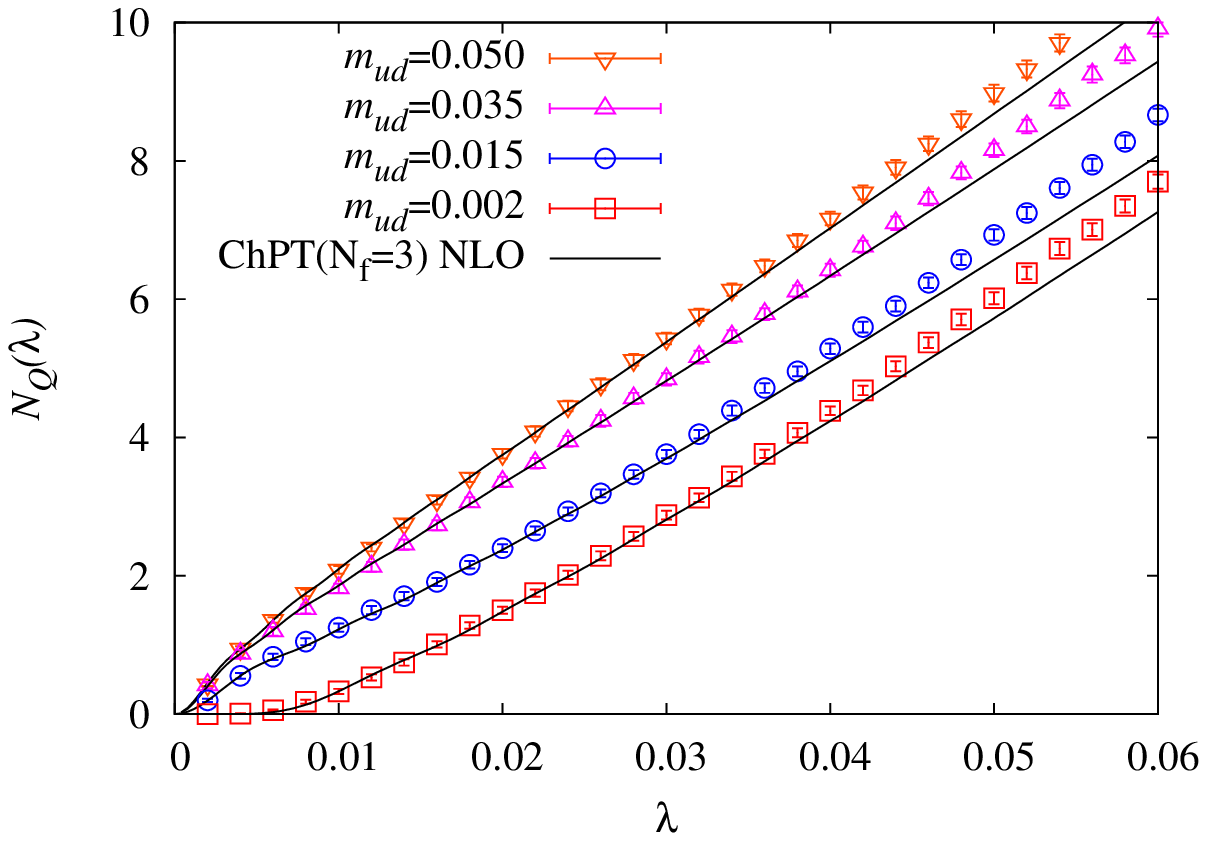}
  \caption{
    Same as Figure~\ref{fig:Nf2} but for the $N_f=2+1$ data at $m_s=0.080$. 
    Solid curves show the NLO ChPT ($N_f=3$) formula.
  }
  \label{fig:Nf3}
\end{figure}

Figures~\ref{fig:Nf2} and \ref{fig:Nf3} compare the spectral density
and its integral at various sea quark masses.
The data are obtained for $N_f=2$ (Figure~\ref{fig:Nf2}) and $N_f=2+1$
(Figure~\ref{fig:Nf3}) with up and down quark masses in the $p$ regime 
($m_{ud}$ = 0.050, 0.035, 0.015) and in the $\epsilon$ regime
($m_{ud}$ = 0.002).
Note that the $\beta$ value of the $\epsilon$ regime run in $N_f=2$
QCD is slightly higher ($\beta=2.35$) than in other runs
($\beta=2.30$).
In the plot, we adjust the value of $\lambda$ by a factor of 1.065
which corresponds to the ratio of lattice spacings between the two
$\beta$ values.

In the plots, we clearly observe the sea quark mass dependence.
For heavier sea quarks, the spectral density shows a higher peak near
the lowest eigenvalue (around $\lambda\sim$ 0.004), and the peak
height becomes lower by reducing the quark mass.
As one enters the $\epsilon$ regime, the lowest eigenvalue is pushed
up to $\lambda\sim$ 0.015.
This is what we expect for the effect of the fermionic
determinant $\prod_k(\lambda_k^2+m_{ud}^2)^2$.
Namely, when the quark mass is reduced to the value around (or below)
the lowest eigenvalue, those eigenvalues are suppressed.

The expectation from NLO ChPT precisely follows the lattice data
(solid curves in Figures~\ref{fig:Nf2} and \ref{fig:Nf3}).
Here the values of $\Sigma_{\rm eff}$ and $F$ are determined for each
set of sea quark mass, so that the comparison is not parameter-free.
But, still the precise agreement of the shape of the spectral density
is encouraging.
We investigate the mass dependence of $\Sigma_{\rm eff}$ in the next
section.

\begin{figure}[tbp]
  \centering
  \includegraphics[width=14cm]{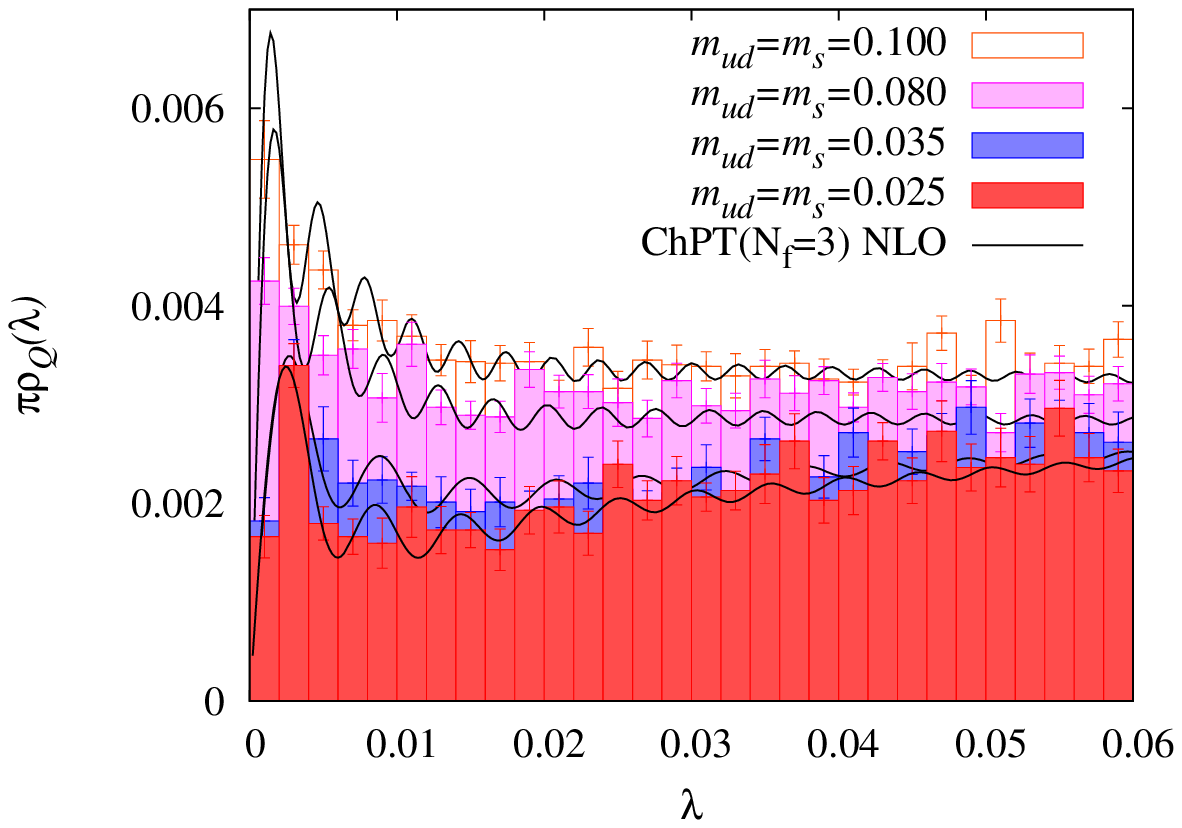}
  \includegraphics[width=14cm]{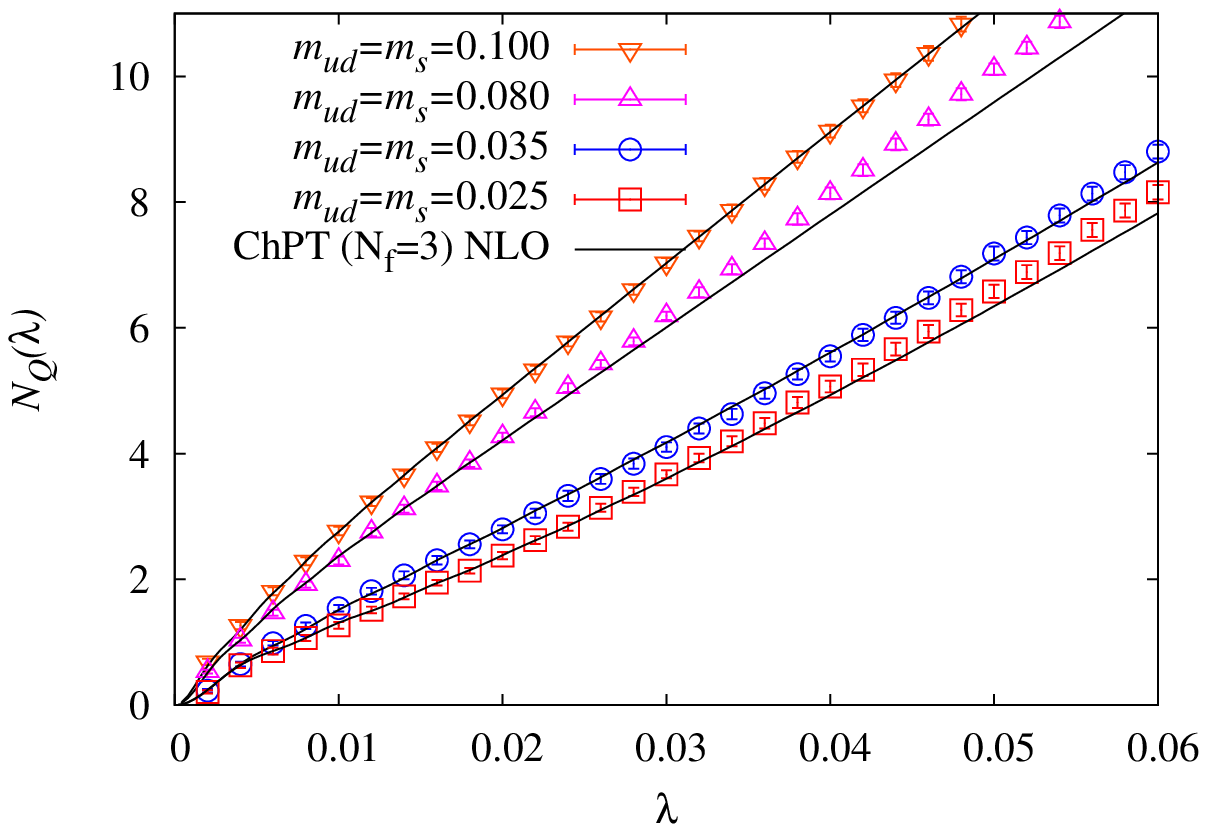}
  \caption{
    Same as Figure~\ref{fig:Nf3} but for degenerate $N_f=3$ data.
  }
  \label{fig:Nf3deg}
\end{figure}

In Figure~\ref{fig:Nf3deg}, 
similar plots are also shown for the $N_f=3$ lattice data where three
sea quarks have a degenerate mass ($m_{ud}=m_s$).
We have four values of the quark mass (0.100, 0.080, 0.035 and 0.025)
in the $p$ regime.
In $N_f=3$ QCD, we observe a stronger dependence on the quark mass
than in $N_f$ = 2 or $2+1$, as suggested by the NLO formula 
for 
$\Sigma_{\rm eff}$ in (\ref{eq:Sigmaeff}).

\subsection{Topological charge}

\begin{figure}[tbp]
  \centering
  \includegraphics[width=14cm]{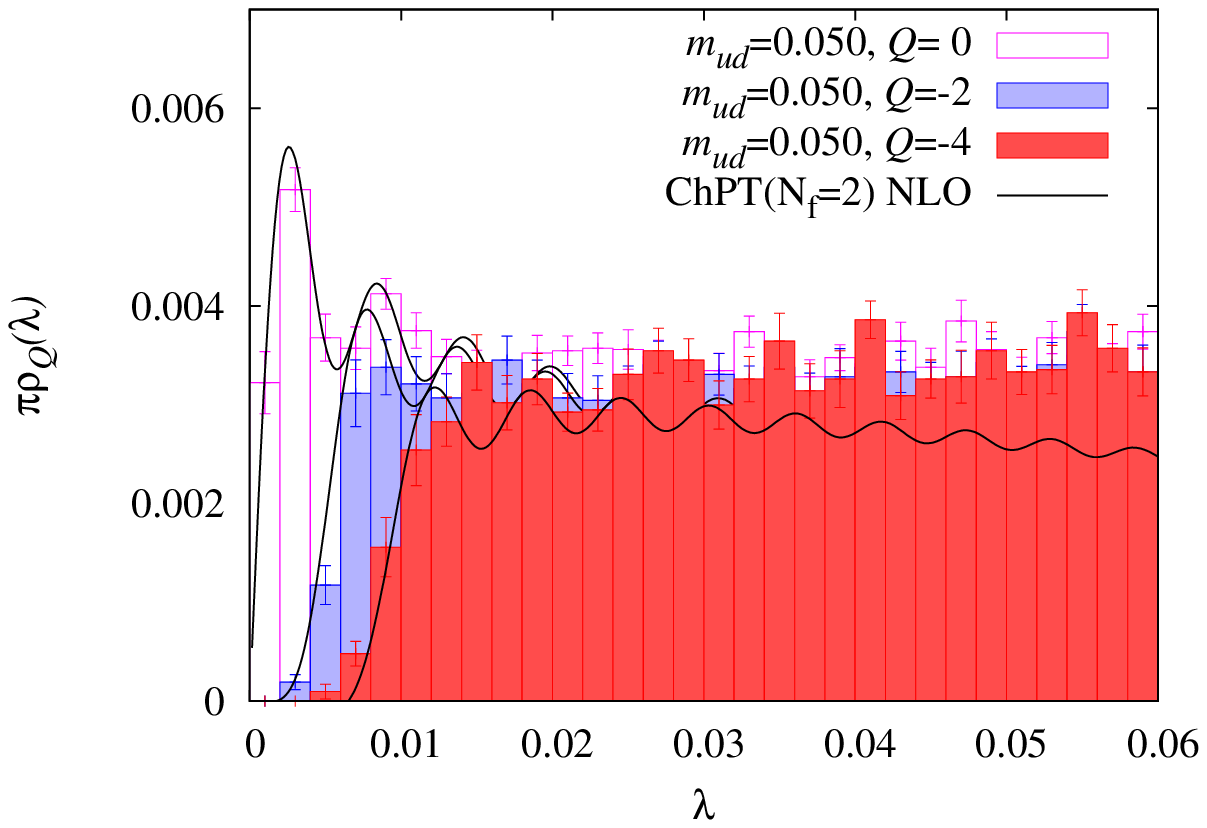}
  \includegraphics[width=14cm]{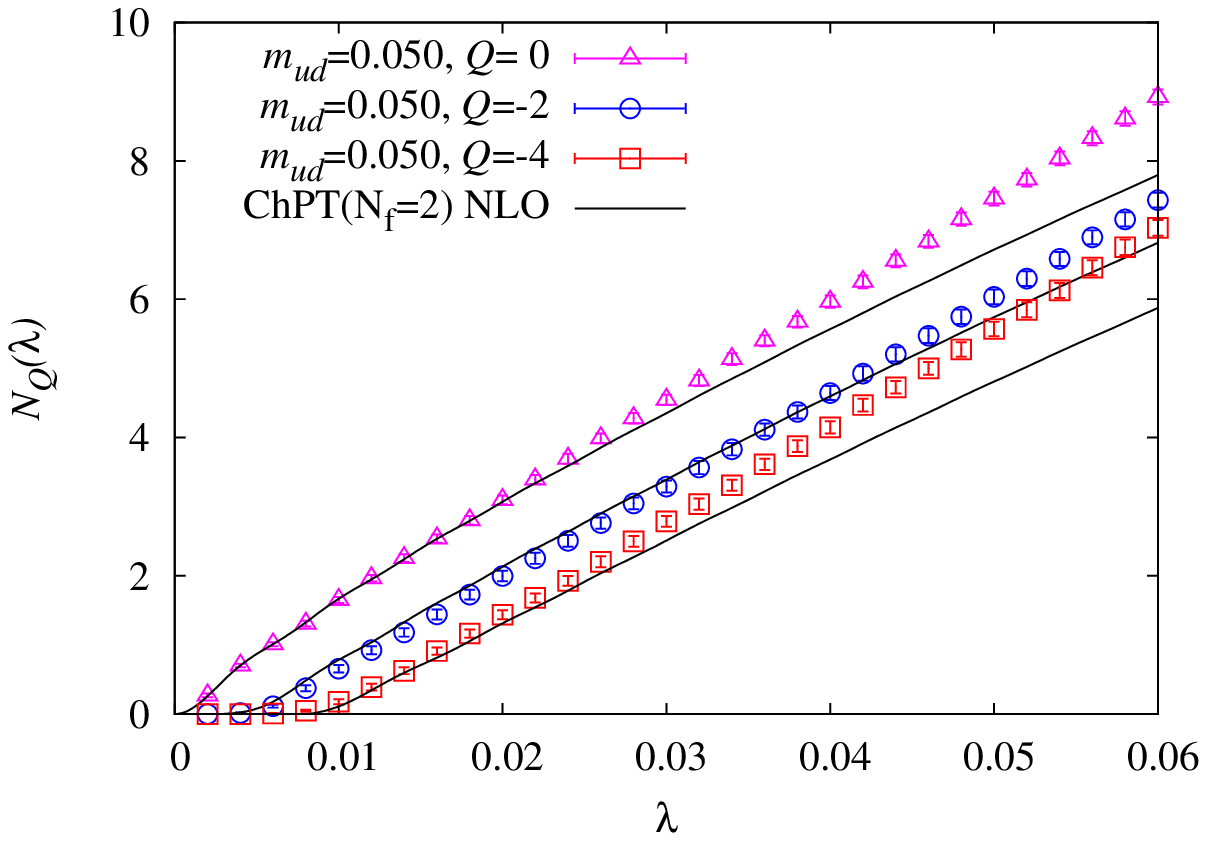}
  \caption{
    Dependence on the topological charge of
    the $N_f=2$ lattice QCD results at $m_{ud}=0.050$.
    For the ChPT curves (solid), the same input values  
    of $\Sigma_{\rm eff}$, $F$ determined from $Q=0$ are used.
  }
  \label{fig:QNf2}
\end{figure}

\begin{figure}[tbp]
  \centering
  \includegraphics[width=14cm]{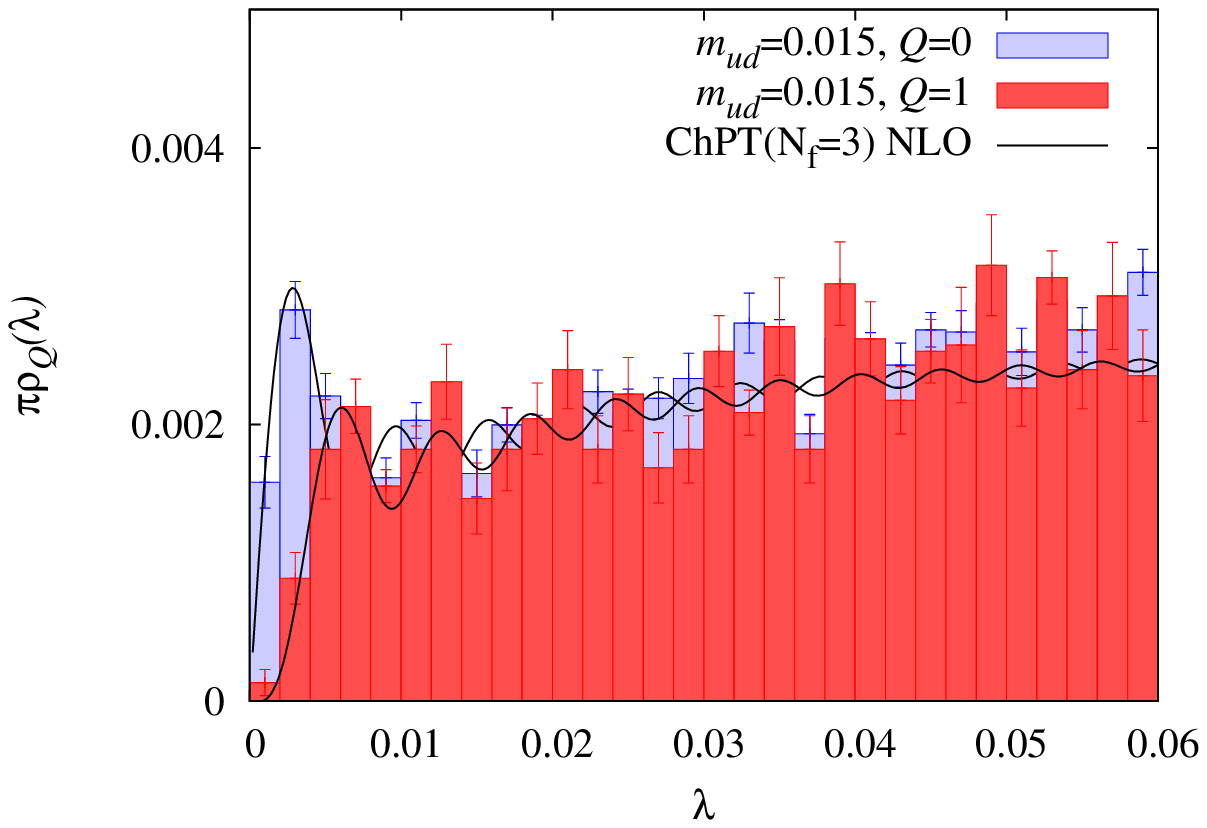}
  \includegraphics[width=14cm]{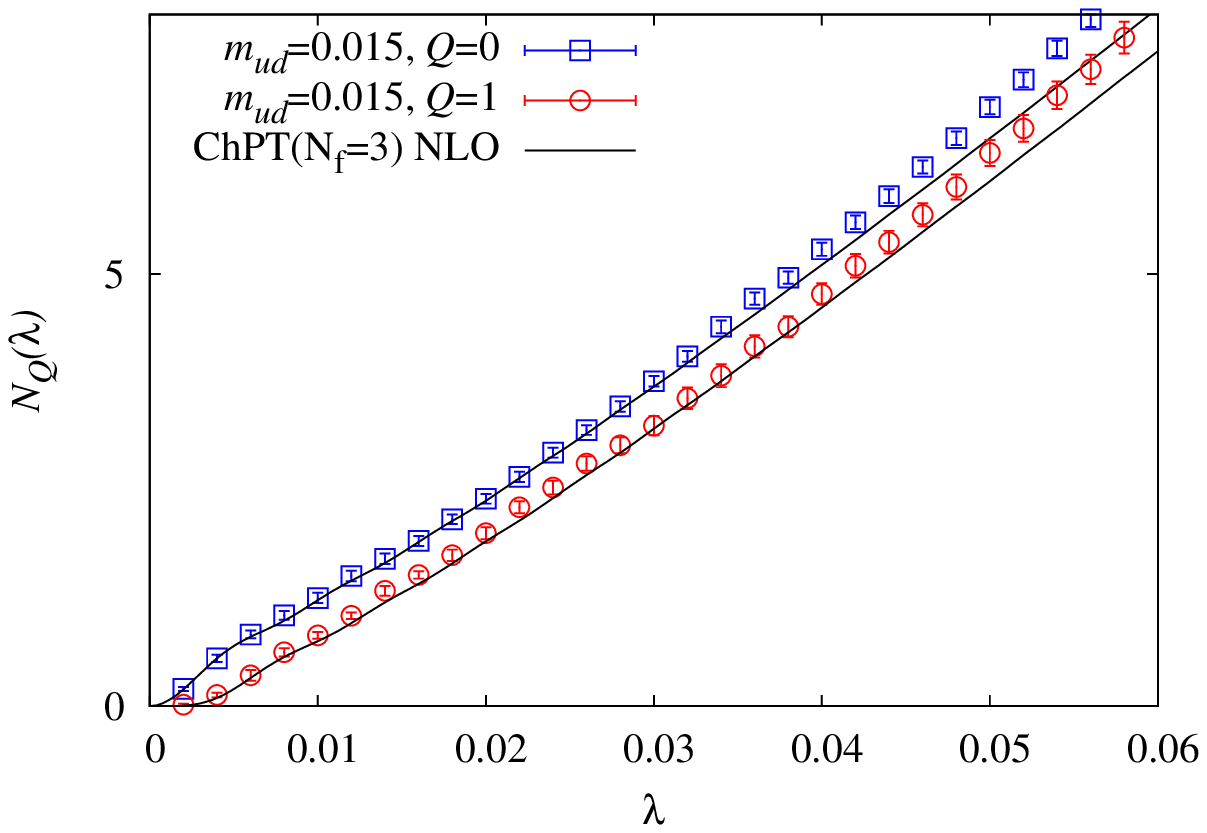}
  \caption{
    Same as Figure~\ref{fig:QNf2} but for $N_f=2+1$ QCD 
    at $m_{ud}=0.015$ and $m_s=0.080$.
  }
  \label{fig:QNf3}
\end{figure}

In the low eigenvalue region, the Dirac spectral density is known to
be sensitive to the topological charge $Q$ of the gauge fields, 
which is clearly seen in Figures~\ref{fig:QNf2} and \ref{fig:QNf3}.
The solid curves in the plots show the expectation from the NLO ChPT
with input parameters determined from the $Q=0$ lattice data.
Therefore, there is no free parameter to be adjusted 
in the curves for the $Q\neq 0$ cases.
We observe that the $Q$ dependence of the lattice data is
qualitatively well described by ChPT.

Note, however, that the extracted values of $\Sigma_{\rm eff}$ and $F$
from $Q\neq 0$ data show a 2.2 $\sigma$ difference for $N_f=2$
while they are consistent in the $N_f=2+1$ case.
Since the topological charge dependence is a part of finite volume
effects \cite{Brower:2003yx,Aoki:2007ka,Aoki:2009mx}, which should be
accounted for by the effective theory analysis, 
the deviation suggests possible higher order effects in the $1/L$
expansion of ChPT. 
Indeed, the $N_f=2$ runs are carried out
with a shorter temporal extent $T=32$ than that of $N_f=3$ ($T=48$),
so that the lattice volume is $\sim 17$\% smaller in the physical
unit. Higher order finite volume effects may therefore be enhanced
for $N_f=2$. In the final results of the $N_f=2$ data, we add 
this $\sim $11\% deviation as an estimate of the systematic error 
due to the finite volume.

\subsection{Finite volume}

\begin{figure}[tbp]
  \centering
  \includegraphics[width=14cm]{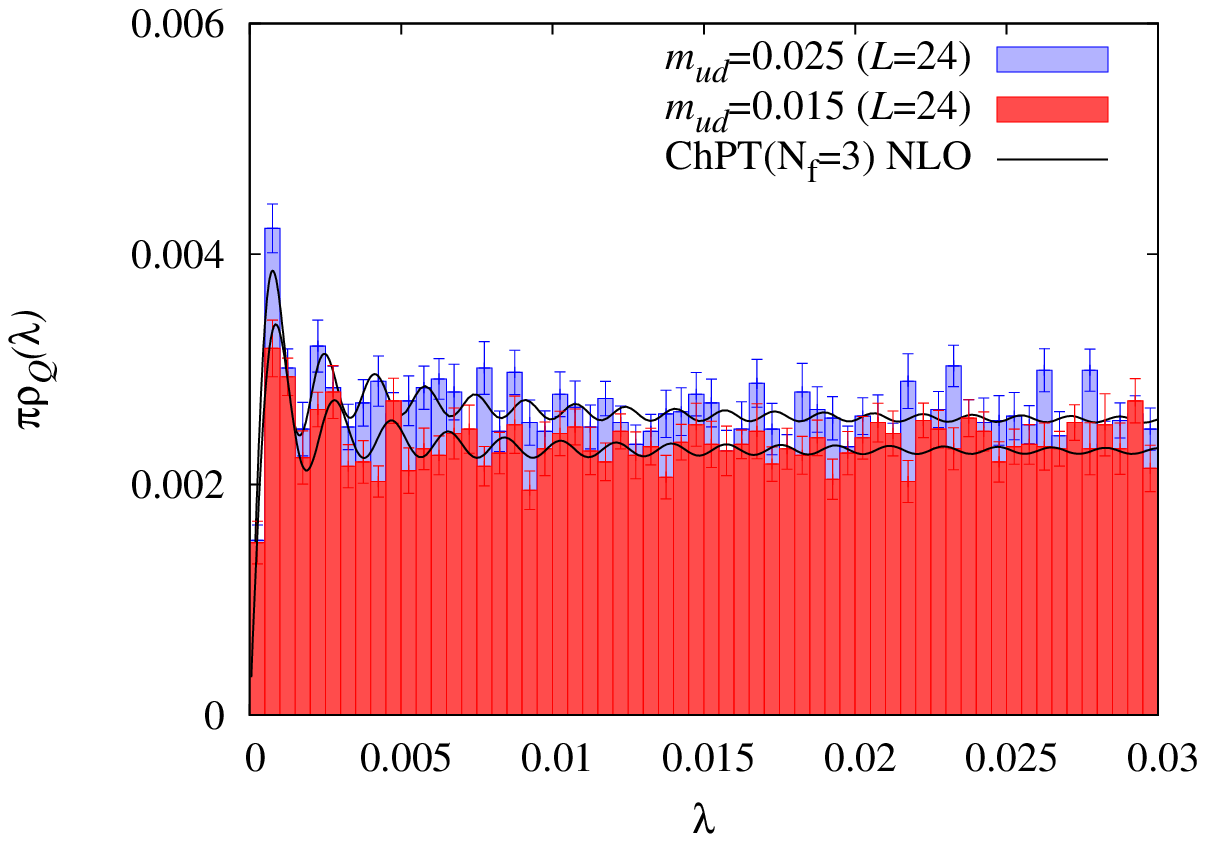}
  \includegraphics[width=14cm]{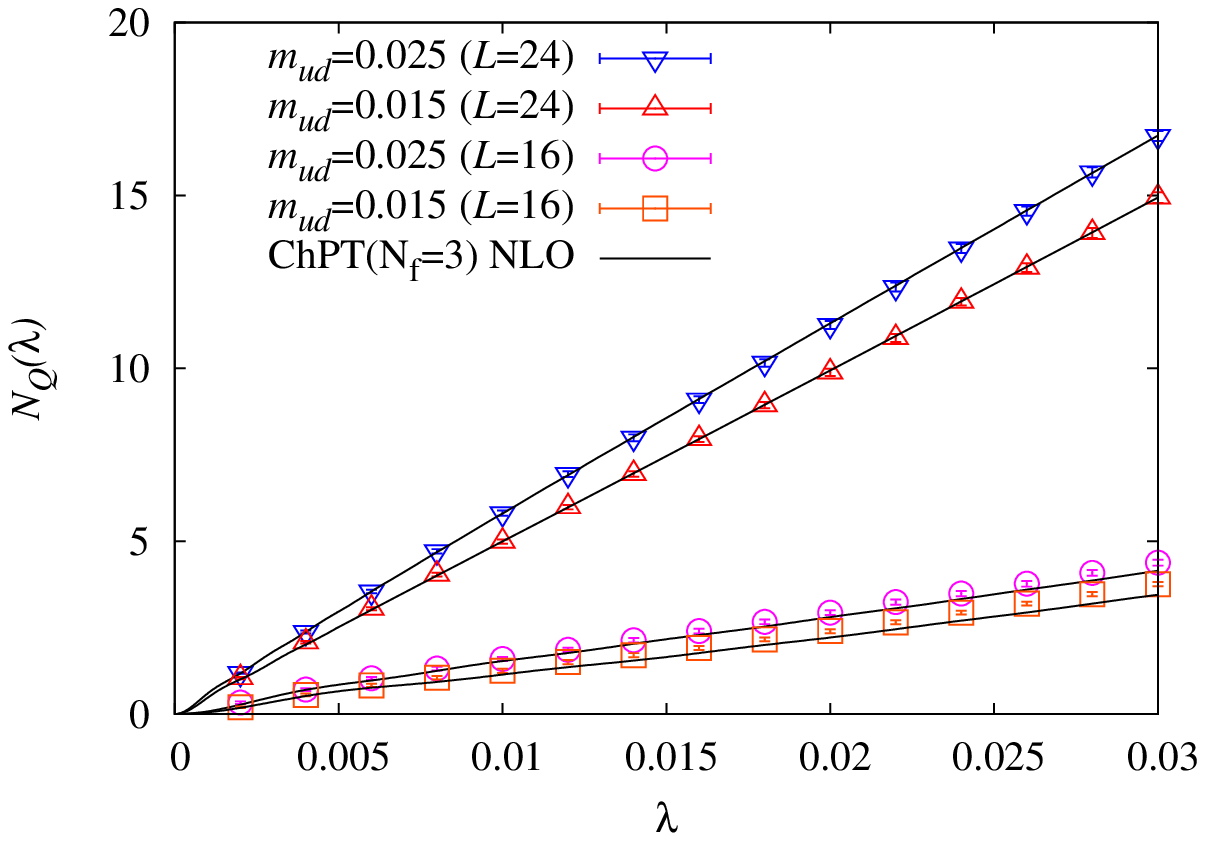}
  \caption{
    Spectral density (top) and the mode number (bottom)
    on the larger volume lattice ($L=24$ [$\sim $2.7 fm]) at $m_{ud}=0.015$ and 0.025
    with a fixed value of $m_s=0.080$.
    For comparison, $L=16$ lattice results as well as the
    ChPT predictions (solid curves) are shown in
    the bottom panel where the same input values of 
    $\Sigma_{\rm eff}$, $F$ (from $L=24$ results) are used.
  }
  \label{fig:V}
\end{figure}

The finite volume scaling can be tested more explicitly with the
$N_f=2+1$ runs, for which the $L=24\;(\sim 2.7\;\mbox{fm})$ 
lattice data are available.
Here we note that the comparison of $\Sigma_{\rm eff}$ obtained 
on the $L=16$ and $L=24$ 
lattices is not straightforward, because
there is a finite volume effect encoded in $\bar{g}_1(M^2)$ in the
definition of $\Sigma_{\rm eff}$ (\ref{eq:Sigmaeff}).
It is still possible to analytically convert the values of
$\bar{g}_1(M^2)$ in different volumes.
The results for $\Sigma_{\rm eff}$ obtained on the $L=24$ lattice are
converted to those at the smaller volume as 
$\Sigma_{\rm eff}$ = $0.00273(06)\to 0.00305(15)$ at $m_{ud}=0.015$
and 
$\Sigma_{\rm eff}$ = $0.00299(06)\to 0.00336(12)$ at $m_{ud}=0.025$,
which agrees with the values calculated on the $L=16$ lattice:
$\Sigma_{\rm eff}$ = 0.00314(18) and 0.00333(18), respectively.
In fact, as shown in Figure~\ref{fig:V}, we find that the same inputs
of (converted) $\Sigma_{\rm eff}$ and $F$ describe the data at
different volumes very well.

From these analysis, the systematic error due to finite volume is
estimated as $\sim 3$\%, which is taken from the difference between the
$L=16$ and $L=24$ (after the conversion) results.


\newpage
\section{Determination of the chiral condensate}
\label{sec:determination}

The extracted values of $\Sigma_{\rm eff}$ and $F$ for each lattice
ensemble are summarized in Table~\ref{tab:SigmaeffFeff}.
Note that $\Sigma_{\rm eff}$ is extracted at the NLO accuracy, 
while the value of $F$ which first appears in the NLO term,
might receive larger systematic corrections from the NNLO
contributions.

As already noted above, 
since $m_s$ is fixed at a large value (0.080 or 0.100) in the
$N_f=2+1$ ensembles, there is little difference between the reduced
$N_f=2$ and $N_f=2+1$ ChPT analysis near the chiral limit of $m_{ud}$: 
$\Sigma_{\rm eff}$ and $F$ are almost equal within the statistical errors.
The difference between $m_s$ = 0.080 and $m_s$ = 0.100
is also small (always less than 1$\sigma$), and therefore
we concentrate on the data at $m_s=0.080$ in the following.

Now we analyze the sea quark mass dependence of $\Sigma_{\rm eff}$.
The NLO formula for $\Sigma_{\rm eff}$ (\ref{eq:Sigmaeff}) contains
the low energy constants $\Sigma$, $F$ and $L_6$ as parameters.
The chiral condensate $\Sigma$, in particular, appears at the leading
order, and its determination through the quark mass dependence of
$\Sigma_{\rm eff}$ is valid at the NLO accuracy, 
while the other parameters controlling the NLO correction can be
determined at the leading-order.

In the fit of the lattice data, we attempt two procedures:
a 3-parameter ($\Sigma$, $F$, $L_6$) fit without any additional inputs,
and 
2-parameter ($\Sigma$, $L_6$) fits for several input values of $F$.  
In the 2-parameter fits, 
the value of $F$ determined in our previous works both in the
$p$ regime ($F$ = 0.0474(30)) \cite{Noaki:2008iy} and in the
$\epsilon$ regime ($F$ = 0.0524(34)) \cite{Fukaya:2007pn} are used for
the $N_f=2$ lattice ensembles.
Any difference due to the different input values of $F$
suggests some systematic error (although it turns out to be only $\sim 1.1\sigma$
in the final result).
For the $N_f$ = $2+1$ and 3 runs, we use a naive (linear) chiral
limit of $F$ given in Table~\ref{tab:SigmaeffFeff}, of which values are
$F=0.0411$ for $N_f=3$ ChPT and $F=0.0406$ 
for reduced $N_f=2$ ChPT, respectively.

\begin{figure}[tbp]
  \centering
  \includegraphics[width=13cm]{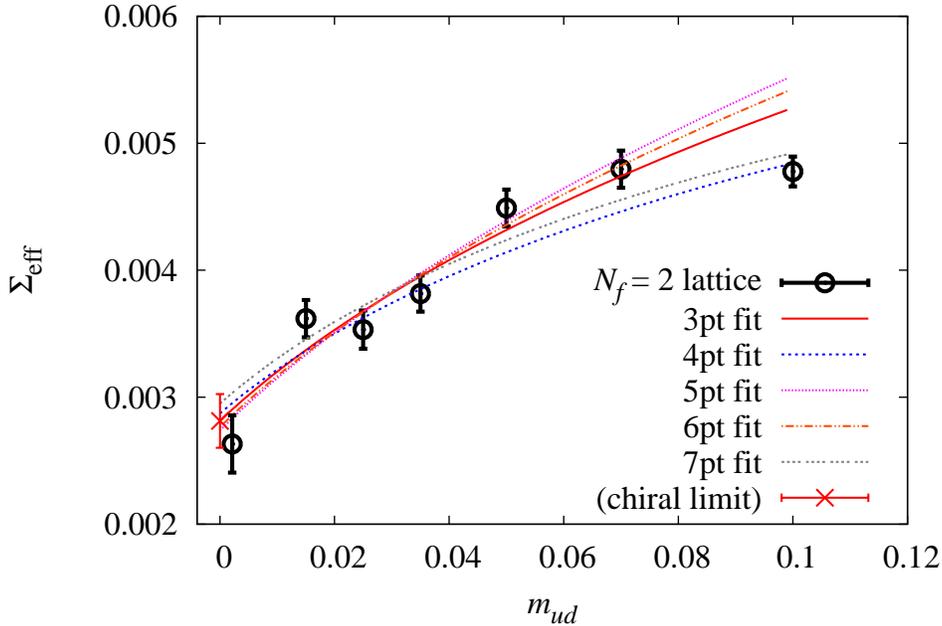}
  \caption{
    Chiral extrapolation of $\Sigma_{\rm eff}$ in $N_f=2$ QCD.
    The data point in the $\epsilon$ regime ($m_{ud}$ = 0.002) is
    rescaled to absorb the small difference of the lattice spacing.
    The two-parameter fits with an input $F=0.0474$ 
    \cite{Noaki:2008iy} for various number of data points included in
    the fit are drawn together with the lattice data points (open
    circles). 
    The chiral limit is that for the 3-point fit.
  }
  \label{fig:SigmaefffitNf2}
\end{figure}

The chiral extrapolation of $\Sigma_{\rm eff}$ in $N_f=2$ QCD is shown
in Figure~\ref{fig:SigmaefffitNf2}.
From the plot, we can see the crucial role played by the
data point at $m_{ud}$ = 0.002 which is in the $\epsilon$ regime.
Without this data point, one may naively expect that the data do not
have enough sensitivity to probe the curvature due to the chiral
logarithm and the chiral extrapolation favors a larger value of
$\Sigma$ ($\sim$ 0.0032).
Taking the $\epsilon$ regime data into account, the presence of chiral
logarithm is consistent with the negative curvature seen in the data.

\begin{table}[tbp]
\centering
\begin{tabular}{ccccc}
\hline\hline
\multirow{2}{*}{$m_{ud}$ fit range} & \multicolumn{3}{c}{$N_f$=2 ChPT LECs} 
&\multirow{2}{*}{$\chi^2/$d.o.f.}\\
 & $\Sigma$ & $F$ & $L^r_6$ & \\ 
\hline
2prm fit &($F=$0.0474)&& \\
0.002-0.025 & 0.00243(18) & [0.0474] & -0.00004(21) & 4.8\\
0.002-0.035 & 0.00246(15) & [0.0474] & -0.00009(13) & 2.5\\
0.002-0.050 & 0.00233(12) & [0.0474] &  0.00007(10) & 2.4\\
0.002-0.070 & 0.00233(09) & [0.0474] &  0.00006(07) & 1.8\\
0.002-0.100 & 0.00246(07) & [0.0474] & -0.00004(03) & 2.4\\
2prm fit &($F=$0.0524)\\
0.002-0.025 & 0.00250(19) & [0.0524] &  0.00012(30) & 5.2\\
0.002-0.035 & 0.00255(15) & [0.0524] &  0.00002(18) & 2.7\\
0.002-0.050 & 0.00243(12) & [0.0524] &  0.00021(14) & 2.4\\
0.002-0.070 & 0.00246(10) & [0.0524] &  0.00018(09) & 1.8\\
0.002-0.100 & 0.00262(08) & [0.0524] &  0.00001(04) & 2.8\\
3prm fit & & & \\
0.002-0.035 & 0.00174(45) & 0.0290(71) & -0.00024(03) & 2.5\\
0.002-0.050 & 0.00246(68) & 0.0547(78) &  0.00038(90)  & 3.5\\
0.002-0.070 & 0.00237(38) & 0.0489(15) &  0.00009(35)  & 2.4\\
0.002-0.100 & 0.00206(26) & 0.0386(48) & -0.00009(02) & 2.3\\
\hline
\end{tabular}
\caption{
  $N_f=2$ lattice results for $\Sigma$, $F$ and
  $L_6^r(\mu_{sub}=\mbox{770~MeV})$ extracted comparing with the $N_f=2$ ChPT
  formula.
  The results for 2- and 3-parameter fits are listed.
  The values in the parenthesis $[\cdots]$ are used as an input of the
  chiral fit.
}
\label{tab:fitNf2}
\end{table}

The extracted values of the LECs in $N_f=2$ ChPT are summarized in
Table~\ref{tab:fitNf2}. 
We attempt the 2-parameter fits of various number of data points,
3--7, taken from the lowest quark mass and the 3-parameter fits with 4--7
data points.
In the table, the range of $m_{ud}$ used in the fit is listed.
For the 2-parameter fits, we take two input values of $F$.
The quality of the fits can be inferred from the value of
$\chi^2/$d.o.f. also listed in the table.

As far as the heaviest point is discarded in the fits, the resulting
value of $\Sigma$ is insensitive to the input value of $F$, and it is
consistent with the three-parameter fit as well.
On the other hand, the determination of $L_6^r$ is unstable, but
all the data suggest $|L_6^r|<0.001$.
We take $\Sigma=0.00246(15)$ and $L_6^r=-0.00009(13)$
as the central values, which are from the 4-point fit with the input
$F=0.0474$. 
For the final results of this paper, we 
take the deviation from the other input value of $F$,
as well as the other fitting ranges,
as a systematic error due to the chiral extrapolation.

\begin{figure}[tbp]
  \centering
  \includegraphics[width=13cm]{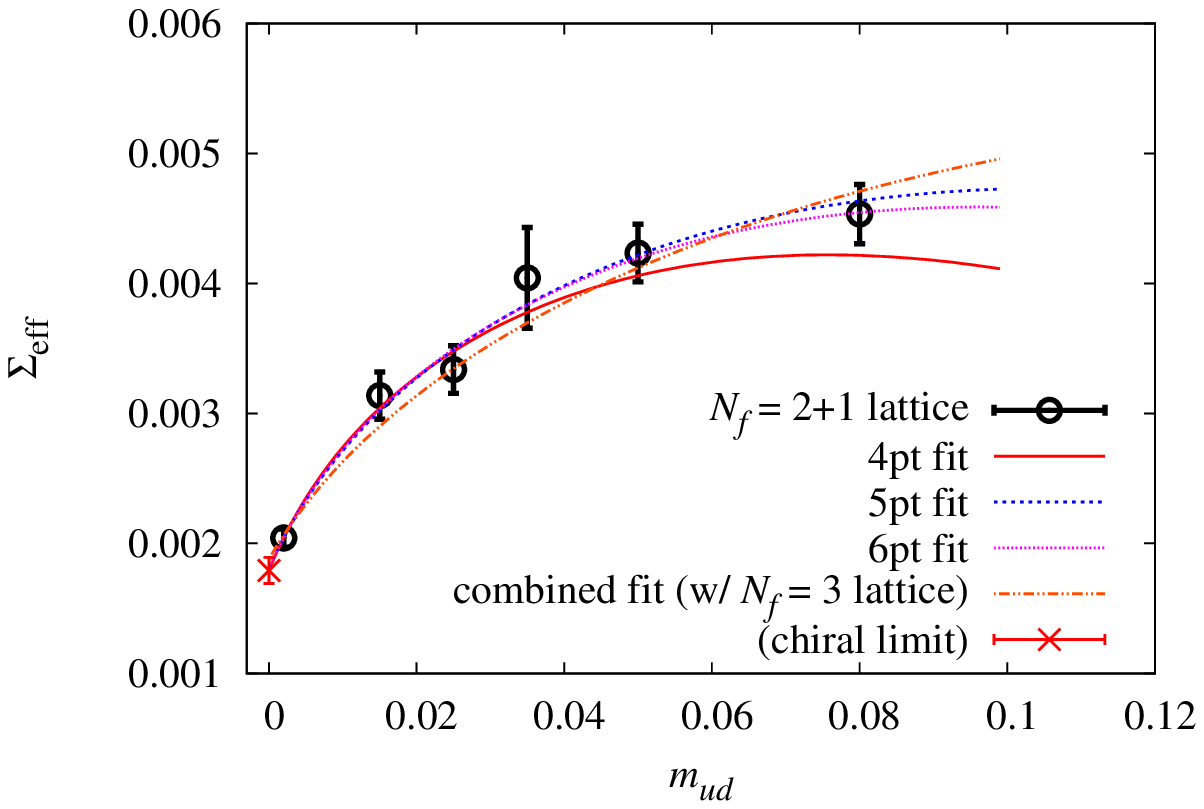}
  \caption{
    Same as Figure~\ref{fig:SigmaefffitNf2} but for $N_f$ = 2+1 QCD.
    The fit curve corresponds to that of the 3-parameter fit.
  }
  \label{fig:SigmaefffitNf3}
\end{figure}

\begin{table}[tbp]
\centering
\begin{tabular}{ccccc}
\hline\hline
\multirow{2}{*}{$m_{ud}$ fit range} & \multicolumn{3}{c}{$N_f$=3 ChPT LECs} 
&\multirow{2}{*}{$\chi^2/$d.o.f.}\\
 & $\Sigma^{\rm phys}$ & $F$ & $L^r_6$ & \\ 
\hline
2prm fit \\
0.002-0.025 & 0.00186(09) & [0.0411] & -0.00013(09) & 1.0\\
0.002-0.035 & 0.00186(09) & [0.0411] & -0.00014(08) & 0.6\\
0.002-0.050 & 0.00186(09) & [0.0411] & -0.00014(07) & 0.5\\
0.002-0.080 & 0.00185(08) & [0.0411] & -0.00014(07) & 0.4\\
3prm fit & & & \\
0.002-0.035 & 0.00185(10) & 0.0433(13) & -0.00023(53) & 1.3\\
0.002-0.050 & 0.00186(09) & 0.0406(05) & -0.00012(25)  & 0.7\\
0.002-0.080 & 0.00186(08) & 0.0413(02) & -0.00015(09)  & 0.5\\
\hline
\end{tabular}
\caption{
  $N_f=2+1$ lattice results for $\Sigma$, $F$ and
  $L_6\;(\mu_{sub}=\mbox{770~MeV})$ extracted using the $N_f=3$ ChPT
  formula. 
  The results for 2- and 3-parameter fits are listed.
  The values with parenthesis $[\cdots]$ are used as an input for the
  chiral fit. 
  $\Sigma^{\rm phys}$ denotes the $m_{ud}=0$ and $V=\infty$ limit
  of $\Sigma_{\rm eff}$ with $m_s=0.080$ fixed.
}
\label{tab:fitNf2+1}
\end{table}

\begin{table}[tbp]
\centering
\begin{tabular}{ccccc}
\hline\hline
\multirow{2}{*}{$m_{ud}$ fit range} & \multicolumn{3}{c}{reduced $N_f$=2 ChPT LECs} 
&\multirow{2}{*}{$\chi^2/$d.o.f.}\\
 & $\Sigma$ & $F$ & $L^r_6$ & \\ 
\hline
2prm fit \\
0.002-0.025 & 0.00199(06) & [0.0406] & -0.00044(10) & 0.9\\
0.002-0.035 & 0.00197(06) & [0.0406] & -0.00040(09) & 0.7\\
0.002-0.050 & 0.00197(05) & [0.0406] & -0.00039(06) & 0.4\\
0.002-0.080 & 0.00198(05) & [0.0406] & -0.00042(04) & 0.5\\
3prm fit & & & \\
0.002-0.035 & 0.00197(09) & 0.0407(10) & -0.00042(51) & 1.3\\
0.002-0.050 & 0.00198(07) & 0.0416(05) & -0.00038(21)  & 0.6\\
0.002-0.080 & 0.00196(07) & 0.0399(03) & -0.00044(07)  & 0.5\\
\hline
\end{tabular}
\caption{
  Same as Table~\ref{tab:fitNf2+1} but extracted using reduced
  $N_f=2$ ChPT.
}
\label{tab:fitreducedNf2}
\end{table}

For the $N_f=2+1$ lattice data, the chiral extrapolation is more
stable because of the precise data point in the $\epsilon$ regime,
which is simply due to higher statistics we accumulated.
Figure~\ref{fig:SigmaefffitNf3} clearly shows the logarithmic curvature.
Namely, a naive linear extrapolation of four data points in the
$p$ regime ($m_{ud}$ = 0.015--0.050) would lead to $\sim$ 0.0028 in the
chiral limit, while the $\epsilon$ regime point is lower
($\sim$ 0.0020).

We analyze the $N_f=2+1$ lattice data listed in Table~\ref{tab:SigmaeffFeff}
using the 2- and 3-parameter fits.
We consistently use the $N_f=3$ formula for the
data obtained with $N_f=3$ ChPT (fourth column in Table~\ref{tab:SigmaeffFeff}).
The same applies for the reduced $N_f=2$ ChPT analysis.
The fit results 
are presented in Table~\ref{tab:fitNf2+1} and
in Table~\ref{tab:fitreducedNf2}.
The curves in Figure~\ref{fig:SigmaefffitNf3} represent the $N_f=3$
fits for various numbers of data points.
We find that all the curves go through the data points except for the
heaviest one.
The chiral limit shown by a cross symbol is almost unchanged by taking
different fit schemes (2-parameter or 3-parameter) and the number of
data points included in the fit (4, 5, or 6).
This is because the precise $\epsilon$ regime point works as an
anchor near the chiral limit.

For $N_f=3$ ChPT, $\Sigma^{\rm phys}$ listed in
Table~\ref{tab:fitNf2+1} denotes the value of $\Sigma_{\rm eff}$ in
the limit of $m_{ud}\to 0$ and $V\to \infty$ with a fixed strange
quark mass $m_s=0.080$. 
This corresponds to the {\it physical} value of the chiral condensate
$-\langle\bar{u}u\rangle=-\langle\bar{d}d\rangle$ defined in the limit
of vanishing up and down quark masses.


From Tables~\ref{tab:fitNf2+1} and \ref{tab:fitreducedNf2}, we can see
that results for $\Sigma^{\rm phys}$ obtained via $N_f=2+1$ and reduced
$N_f=2$ ChPT formulas are consistent with each other.
(We assume that $\Sigma^{\rm phys}$ in the $N_f=2+1$ theory corresponds to
$\Sigma$ in reduced $N_f=2$ ChPT.) 
The result is also stable against the changes of 
the number of fitting parameters and the fitting range.
For the other parameters, $F$ and $L^r_6$, we find larger dependence on
the choice of fit procedures, which is expected because they only
appear in the NLO terms.

We take $\Sigma^{\rm phys}=0.00186(09)$, $F=0.0406(05)$ and $L_6^r=-0.00012(25)$
as the central values, which are obtained from the 5-point fit with
three free parameters in $N_f=3$ ChPT. 
As mentioned above, the other results are used to estimate systematic
errors. 

\begin{figure}[tbp]
  \centering
  \includegraphics[width=13cm]{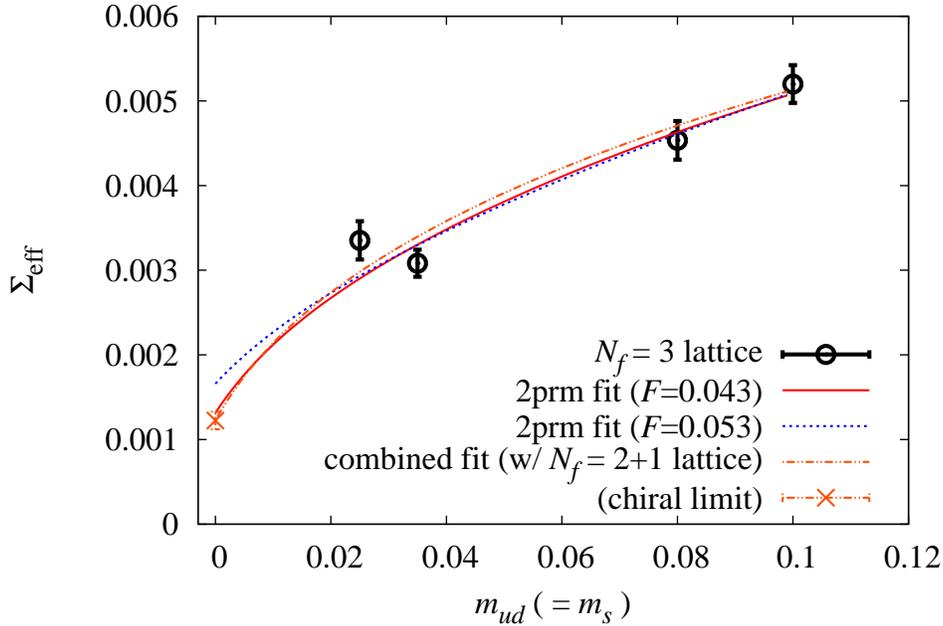}
  \caption{
    Same as Figure~\ref{fig:SigmaefffitNf2} but for the degenerate
    ($m_{ud}=m_s$) $N_f=3$ QCD.
    Results of the 2-parameter fits and a combined $N_f=3$ fit are
    plotted.
  }
  \label{fig:SigmaefffitNf3deg}
\end{figure}

\begin{table}[tbp]
\centering
\begin{tabular}{cccccc}
\hline\hline
\multirow{2}{*}{$m_{ud}, m_s$ fit range} & \multicolumn{4}{c}{$N_f$=3 ChPT LECs} 
&\multirow{2}{*}{$\chi^2/$d.o.f.}\\
 & $\Sigma$ & $\Sigma^{\rm phys}$ & $F$ & $L^r_6$ & \\ 
\hline
2prm fit &&& \\
0.025-0.100 & 0.00152(11) & 0.00204(04)  & [0.0431] & 0.00010(10) & 3.1\\
0.025-0.100 & 0.00182(14) & 0.00242(06)  & [0.0531] & 0.00031(17) & 2.8\\
3prm fit & \multicolumn{3}{c}{(combined with $N_f=2$+1 data)} \\
0.002-0.100 & 0.00145(12) & 0.00191(08) & 0.0401(17) & 0.00003(7) & 1.7\\
\hline
\end{tabular}
\caption{
The degenerate $N_f=3$ lattice results for the LECs of $N_f=3$ ChPT.
Here, $\Sigma$ denotes the chiral condensate at 
the limit $m_{ud}=m_s=0$ while $\Sigma^{\rm phys}$ is the one
with $m_s=0.080$ fixed.
}
\label{tab:fitNf3}
\end{table}

For the degenerate ($m_{ud}=m_s$) $N_f=3$ lattice results,
the number of data points is limited to four.
The 3-parameter fit, therefore, does not work and we restrict
ourselves to the 2-parameter fit.
We attempt the fit with two values of $F$, 0.0431 and 0.0531, as the
input.
The former value is an $N_f=3$ chiral limit of $F$ in
Table~\ref{tab:SigmaeffFeff} taken with a linear function in the quark mass
and the latter is the one at the lightest sea quark mass $m_{ud}=m_s=0.025$.
Due to the lack of the $\epsilon$ regime data point, the chiral
limit is not as stable as in the $N_f=2$ or $N_f=2+1$ data.
In fact, the resulting value of $\Sigma$ strongly depends on the input
value of $F$.
Between the two representative values of $F$, $\Sigma$ changes about
20\%. 
Since $F$ controls the NLO effects as seen in (\ref{eq:Sigmaeff}),
this change suggests that the one-loop calculation is
not sufficient to control the $m_{ud}=m_s$ dependence.

We also attempt a combined fit of all the $N_f=2+1$ and $N_f=3$ data
points using the $N_f=3$ ChPT formula.
The total nine data points are simultaneously fitted
with a reasonable $\chi^2/{\rm d.o.f.}$ ($\simeq$ 1.7).
The result is given in Table~\ref{tab:fitNf3} and plotted in
Figure~\ref{fig:SigmaefffitNf3deg} (and \ref{fig:SigmaefffitNf3}) 
which indicate
a strong effect of chiral logarithm in the $N_f=3$ data 
compared to the case of $N_f=2$ shown in
Figure~\ref{fig:SigmaefffitNf2}. 
The chiral limit is less than half of the value at the lowest
$p$ regime point ($m_{ud}=m_s=0.025$).
One should note, however, that the data points
in the fit includes those with $m_s=0.080$ and 0.100
which are out of the typical convergence region of 
chiral expansion $m_i<m_s^{\rm phys}/2$, so that
the result may contain large systematic effects.

It is still remarkable that all the results suggest that 
the chiral condensate $\Sigma$ of
$N_f=3$ QCD is smaller than $\Sigma^{\rm phys}$ of $N_f=2+1$ or 
$\Sigma$ of $N_f=2$ QCD.
This is consistent with the view that the chiral condensate decreases
and eventually disappears as the number of flavor increases and the
asymptotic freedom is lost.
We take $\Sigma=0.00145(12)$, $F=0.0401(17)$ and $L_6^r=0.00003(7)$
determined from the combined fit as the central values of $N_f=3$ ChPT
parameter, 
and will include a $\sim$ 26\% deviation from the 
2-parameter fit with $F=0.0531$ as a systematic error in the final
results.  

So far we have treated the strange quark mass fixed at $m_s=0.080$ in
the $N_f=2+1$ studies. 
In fact, the combined fit of the degenerate $N_f=3$ and 2+1 results
implies that $\Sigma_{\rm eff}$ in the chiral limit of $m_{ud}$ 
changes by less than 1\% when $m_s$ varies between 0.060 and 0.100.
We can, therefore, safely ignore the error due to a slight mismatch of
the strange quark mass from its physical value.
This weak sensitivity to $m_s$ supports the assumption that the 
strange quark at the physical mass 
is almost decoupled from the low energy theory and the
use of the reduced $N_f=2$ ChPT formula to fit the $N_f=2+1$ lattice
QCD data.

\section{Summary and Conclusion}
\label{sec:conclusion}

Before quoting the final results, let us discuss the possible
systematic errors. 

Our simulation uses an exactly chiral-symmetric Dirac operator 
and has reached almost the chiral limit: 
$m_{ud}$=0.002 for the $N_f=2$ and 2+1 runs,
which corresponds to $\sim$ 3~MeV in the physical unit.
The NLO ChPT formula (\ref{eq:rho}) is valid in both the
$\epsilon$ and $p$ regimes.
As a result, the chiral extrapolation of $\Sigma_{\rm eff}$ 
in $N_f=2$ and 2+1 QCD 
is stable.
As discussed in the previous section, by varying the fit range 
(and $N_f$ in the ChPT formula for the $N_f=2+1$ analysis),
we estimate the systematic effects due to the chiral extrapolation
as $(+3.7\%/\!-5.2\%)$ for $\Sigma$ of $N_f=2$ QCD
and $(+6.5\%/\!-0.5\%)$ for $\Sigma^{\rm phys}$ of $N_f=2+1$ QCD,
respectively. 
The upper and lower limits come from the variation of $\Sigma$ with
the various fit schemes.
We also found that the $m_s$ dependence is negligible
in the range $m_s$ = 0.060--0.100, so that
$\Sigma^{\rm phys}$ can be treated as the chiral condensate at the
physical value of the strange quark mass. 

On the other hand, for the degenerate $N_f=3$ lattice data,
our estimate for the systematic error in the chiral fit
is larger because of the bad convergence of $N_f=3$ ChPT
and smaller number of data points.
In fact, we observe that $\Sigma$ moves as large as 26\% depending
on the fit schemes, from which we estimate the systematic error from
this source to be $\pm$26\% for $\Sigma$ in the $N_f=3$ chiral limit.

The finite volume effects have also been discussed in the previous
sections. 
For the $N_f=2$ lattice results, we expect that a possible higher
order effect in the $1/L$ expansion beyond one-loop ChPT is partly
reflected in the difference among different topological sectors.
We thus estimate the systematic error from this source to be
$\pm$11\%.
For the $N_f=2+1$ (and 3) case, we use more direct comparison with
$L=24$ lattice results and the systematic error is estimated as
$(+0.9\%/\!-2.9\%)$.
With a naive order counting, the leading finite volume effect is
estimated as $1/(F^4V)$, which is the two-loop effect in the $p$
expansion.
With $F$ = 71~MeV (see below), this gives a large value ($\sim$ 0.52)
as the size of the two-loop correction.
In fact, including the numerical coefficient $\beta_1$ given in the
Appendix~\ref{app:beta} the one-loop correction is 
$\beta_1/(F^2V^{1/2})\sim$ $-$0.06.
If we assume that the numerical coefficient is also small 
($\sim$ 0.05) at the two-loop order, this naive estimate gives a 3\%
effect, which is in the same ball park as the estimate given above.
The small numerical coefficient at the two-loop level is indeed
obtained in a recent study \cite{Lehner:2010mv}.

Since our lattice studies are done at only one value of $\beta$,
it is difficult to estimate the size of discretization effects.
It should be partly reflected in the mismatch of the
observables measured in different ways. 
For instance, the inverse lattice spacing determined
from the $\phi$-meson mass, $1/a=1.774$(17)~GeV,
is 1\% larger than the determination from the $\Omega$-baryon mass,
$1/a=1.759$(10)~GeV \cite{JLQCD:prep}.
The latter corresponds to the Sommer scale $r_0$ = 0.51~fm, which is
higher than the nominal value 0.49~fm or the recently favaored value
0.46--0.47~fm by about 4--10\%. On the other hand, a naive order
counting $(a\Lambda_{\rm QCD})^2$ with $\Lambda_{\rm QCD}\sim 450$~MeV
suggests a systematic effect of $\sim 7$\%, which is consistent with
the above mismatch. We therefore add this naive estimate, $\pm 7\%$,
as the systematic error due to finite lattice spacing.

We convert the value of the condensate
to the definition in the standard renormalization scheme,
{\it i.e.} the $\overline{\mathrm{MS}}$ scheme.
By using the nonperturbative renormalization technique
through the RI/MOM scheme we obtained the $Z$ factor in our previous
work \cite{Noaki:2009xi} as 
$1/Z_S(2 \mbox{GeV})=0.804(10)(^{+25}_{-33})$ for $N_f=2$
and $0.792(10)(^{+24}_{-26})$ for $N_f=2+1$ and 3,
where the errors are statistical and systematic, respectively\footnote{
The value for $N_f=2+1$ is slightly changed from \cite{Noaki:2009xi}
due to the different determination of the lattice scale,
that affects the renormalization group running of the $Z$ factor.
}.
\begin{table}[tbp]
  \centering
  \begin{tabular}{lcccc}
    \hline
    \hline
    & $N_f=2$ ChPT & \multicolumn{3}{c}{$N_f=3$ ChPT} \\
    & $[\Sigma]^{1/3}$ & $[\Sigma^{\rm phys}]^{1/3}$ & $[\Sigma]^{1/3}$ & $F$ \\
\hline
    renormalization & $^{+1.4}_{-1.1}$ \% 
    & $^{+1.2}_{-1.1}$ \% & $^{+1.2}_{-1.1}$ \% &--\\
    chiral fit & $^{+1.2}_{-1.8}$ \%& $^{+2.2}_{-0.2}$ \% & $\pm 8.7$ \%
    & $\pm 8.0$ \%\\
    finite volume &$\pm 3.7$\% &$^{+0.3}_{-1.0}$\% &$^{+0.3}_{-1.0}$\% 
    & $\pm 3.0$ \%\\
    finite $a$ & $\pm 7.0$ \% & $\pm 7.0$ \% & $\pm 7.0$ \%& $\pm 7.0$ \%\\
    \hline
    total & $^{+8.1}_{-8.2}$ \% & $^{+7.4}_{-7.2}$ \% & $\pm 11$ \% &$\pm 11$ \% \\
    \hline
  \end{tabular}
  \caption{Systematic errors for
    $[\Sigma^{\overline{\mathrm{MS}}}(2\mbox{~GeV})]^{1/3}$ and $F$.
    The total errors are 
    obtained by adding each estimate by quadrature.
  } 
  \label{tab:sys}
\end{table}


Including all the systematic effects above,
which are summarized in Table~\ref{tab:sys},
we obtain the chiral condensate of up and down quarks
in their massless limit as
\begin{eqnarray}
\label{eq:SigmaMS}
\Sigma^{\overline{\mathrm{MS}}}(\mathrm{2~GeV}) &=&
\left\{ 
\begin{array}{ll}
[242(05)(20)~\mbox{MeV}]^3 &\mbox{for $N_f=2$ QCD}\\
{}[234(04)(17)~\mbox{MeV}]^3 &\mbox{for $N_f=2$+1 QCD (at physical $m_s$)}\\
{}[214(06)(24)~\mbox{MeV}]^3 &\mbox{for $N_f=3$ QCD (at $m_s=0$)}\\
\end{array}
\right.,
\end{eqnarray}
where the errors are again statistical and systematic, respectively.
Here, the total systematic errors are obtained by
adding each estimate by quadrature.
Note that the result for $N_f=2+1$ is slightly changed from \cite{Fukaya:2009fh}
because a different input for the scale determination is used.
We find a nontrivial $m_s$ dependence on the chiral condensate:
as the strange quark mass goes down from $m_s=\infty$ 
($N_f=2$ QCD) to the chiral limit $m_s=0$,
the value of $\Sigma$ decreases.

The chiral condensate determined in this work (\ref{eq:SigmaMS}) is
consistent with those in our previous results obtained
from the pseudoscalar meson mass \cite{Noaki:2008iy,JLQCD:2009sk}
and from the topological susceptibility 
\cite{Aoki:2007pw,Chiu:2008kt,Chiu:2008jq}.
A similar work done using the $N_f=2$ Wilson fermion and the $p$ regime ChPT
\cite{Giusti:2008vb} quoted
$\Sigma^{\overline{\mathrm{MS}}}(\mathrm{2~GeV})
=276(3)(4)(5)$~MeV, which is slightly higher than our result.
A recent work \cite{Bernardoni:2010nf} in a mixed action approach
(overlap valence + Wilson sea) has also reported a larger value.
More detailed study would be necessary to understand the source of the
discrepancy, if it is significant.

From the NLO terms, we also obtain
\begin{eqnarray}
F &=& 71(3)(8) \mbox{\ MeV} \;\;\;\mbox{for $N_f=3$ QCD}, 
\end{eqnarray}
(or $\sqrt{2}F=$100(4)(11) MeV) and
\begin{eqnarray}
L_6^r(770\mbox{MeV}) &=&
\left\{ 
\begin{array}{ll}
-0.00009(13)(30) &\mbox{for $N_f=2$ QCD},\\
\;\;\;0.00003(07)(17) &\mbox{for $N_f=3$ QCD},\\
\end{array}
\right.
\end{eqnarray}
where the systematic errors are estimated 
in a similar manner.
For $F$, their estimates are listed in Table~\ref{tab:sys}
while for $L_6^r$, the systematic error is dominated by
the one from chiral extrapolation, as seen in
Table~\ref{tab:fitNf2} and~\ref{tab:fitNf2+1}. 
Although the accuracy for these quantities is not as good as
that of the chiral condensate, 
they provide important consistency checks. 

In this study, we have investigated the eigenvalue spectrum of the QCD
Dirac operator, which is free from ultraviolet power divergences.
The lattice QCD results show a good agreement with the ChPT
calculation at NLO in the region of $\lambda$ less than $m^{\rm phys}_s/2$.
In particular, the effect of pion-loop, or the chiral logarithm, is 
clearly seen.
The dependence on the volume $V$, the quark masses $m_{ud}$ and $m_s$,
and the topological charge $Q$ is also well described by ChPT.
Result for the chiral condensate extracted from this study is
therefore robust, as the systematic errors are controlled
except for that coming from the discretization effect.

Our work has also addressed a nontrivial flavor dependence of the
chiral condensate. 
%
As the strange quark mass is reduced towards the
chiral limit, its dynamical effect is seen as lowering 
the value of $\Sigma$.

\begin{acknowledgments}
  HF thanks P.~H.~Damgaard for useful discussions.
  Numerical simulations are performed on the IBM System Blue Gene
  Solution at High Energy Accelerator Research Organization
  (KEK) under a support of its Large Scale Simulation
  Program (No. 09-05). 
  This work is supported in part by the Grant-in-Aid of the
  Japanese Ministry of Education 
  (Nos.~20340047, 20105001, 20105002, 20105003, 20105005, 21684013, 22011012, 22740183),
  National Science Council (Nos. NSC96-2112-M-002-020-MY3, NSC99-2112-M-002-012-MY3)
and NTU-CQSE (Nos. 99R80869, 99R80873).
\end{acknowledgments}

\appendix
\section{Shape coefficient $\beta_n$}
\label{app:beta}

In (\ref{eq:g1nume}), we need to calculate the coefficients $\beta_n$'s 
\cite{Hasenfratz:1989pk}, which depend only on the shape of the
four-dimensional box. 
The definition of $\beta_n$ is given by
\begin{eqnarray}
\beta_n &\equiv  & \left(\frac{-1}{4\pi}\right)^n \left(
\alpha_n+\frac{2}{n(n-2)}\right)\;\;\; (n\neq 2),\;\;\;\;\;
\beta_2 \equiv \frac{\alpha_2 - \ln 4\pi + \gamma - 3/2}{16\pi^2},\\
\alpha_n &\equiv & 
\int_0^1 dt\left\{t^{n-3}\left(S\left(\frac{L^2}{V^{1/2}t}\right)^3
S\left(\frac{T^2}{V^{1/2}t}\right)-1\right)
\right.\nonumber\\&&\left.\hspace{1in}
+
t^{-n-1}\left(S\left(\frac{V^{1/2}}{L^2t}\right)^3
S\left(\frac{V^{1/2}}{T^2t}\right)-1\right)\right\},\\
S(x) &\equiv & \sum^\infty_{k=-\infty} \exp (-\pi k^2 x),
\end{eqnarray}
where $\gamma\sim 0.577215665\cdots$ is the Euler's constant.
Here the summation in $S(x)$ is well approximated by a truncation 
$|k|\le 20$.
For the case with $T/L=$1,2 and 3, the numerical values of $\beta_n$ 
are listed in Table \ref{tab:betan}.

\begin{table}[tbp]
  \centering
  \begin{tabular}{ccccccccccc}
    \hline\hline
    $T/L$ & $\beta_1$ & $\beta_2$ &$\beta_3$ &$\beta_4$ &$\beta_5$ 
    &$\beta_6$ \\
    \hline
    1 & 0.1405 & -2.030$\times 10^{-2}$ & -4.820$\times 10^{-4}$ 
      & 2.531$\times 10^{-5}$ &  -2.238$\times 10^{-6}$ 
      & 2.672$\times 10^{-7}$ \\
    2 & 0.08360 & -1.295$\times 10^{-2}$ & -1.778$\times 10^{-3}$ 
    & 3.265$\times 10^{-4}$ & -9.120$\times 10^{-5}$ &
      3.250$\times 10^{-5}$\\
    3 &   -0.04194 & 0.01215 & -9.508$\times 10^{-3}$ 
    & 3.622$\times 10^{-3}$ & -1.898$\times 10^{-3}$ 
    & 1.248$\times 10^{-3}$\\
    \hline
  \end{tabular}
  \caption{
    Numerical results for $\beta_n$ for $T/L=$ 1, 2 and 3.
  }
  \label{tab:betan}
\end{table} 


\end{document}